\documentclass[pre,twocolumn,superscriptaddress]{revtex4-1}

\usepackage{amsmath}
\usepackage[xdvi]{graphicx}%
\usepackage{dcolumn}
\usepackage{amsmath}
\usepackage{color}
\usepackage{bm}
\usepackage{subfigure}

\usepackage{array}

\usepackage[utf8]{inputenc}
\usepackage[T1]{fontenc}
\usepackage{ae,aecompl}

\usepackage{color}
\definecolor{darkblue}{rgb}{0,0,0.6}
\definecolor{darkred}{rgb}{0.6,0,0}
\definecolor{darkgreen}{rgb}{0,0.6,0}

\usepackage[colorlinks=true,urlcolor=darkblue,citecolor=darkblue,linkcolor=darkred,hyperfootnotes=false]{hyperref}

\newcommand{\Reyc}{{\rm Re}_{\rm c}}
\newcommand{\Rey}{{\rm Re}}

\begin{document}

\title{Method to measure efficiently rare fluctuations of turbulence intensity for turbulent-laminar transitions in pipe flows
}

\begin{abstract}

The fluctuations of turbulence intensity in a pipe flow around the critical Reynolds number is difficult to study but important because they are related to turbulent-laminar transitions. We here propose a rare-event sampling method to study such fluctuations in order to measure the time-scale of the transition efficiently. The method is composed of two parts: (i) the measurement of typical fluctuations (the bulk part of an accumulative probability function) and (ii) the measurement of rare fluctuations (the tail part of the probability function) by employing dynamics where a feedback control of the Reynolds number is implemented. We apply this method to a chaotic model of turbulent puffs proposed by Barkley and confirm that the time-scale of turbulence decay increases super-exponentially even for high Reynolds numbers up to Re = 2500, where getting enough statistics by brute-force calculations is difficult. The method uses a simple procedure of changing Reynolds number that can be applied even to experiments. 
\end{abstract}

\author{Takahiro Nemoto}
\affiliation{Philippe Meyer Institute for Theoretical Physics, Physics Department, \'Ecole Normale Sup\'erieure \& PSL Research University, 24 rue Lhomond, 75231 Paris Cedex 05, France}
\author{Alexandros Alexakis}
\affiliation{Laboratoire de Physique Statistique, \'Ecole Normale Sup\'erieure, CNRS, Universit\'e Pierre et Mari\'e Curie, Universit\'e Paris Diderot, 24 rue Lhomond, 75005 Paris Cedex 05, France}

\date{\today}

\maketitle

\section{Introduction} 

In 1883, Osborne Reynolds used a dimensionless quantity to characterize pipe flows, the well-known Reynolds number $\rm Re$  \cite{reynolds1883experimental}. 
This number, defined from the velocity, density, pipe diameter and the viscosity of fluid,   determines the pattern of flows: the flows tend to be laminar when this number is small and tend to be turbulent when it is large. Reynolds himself believed that there is a transition value $\Reyc$, so-called critical Reynolds number, that distinguishes these two patterns of flows. After his proposition, however, many experiments and numerical simulations revealed that the problem was more complex than expected \cite{darbyshire_mullin_1995,faisst_eckhardt_2004,Eckhardt449}. First of all, linear stability analysis shows that the laminar flows are stable for any Reynolds number if the perturbation added to the pipe is infinitesimally small. This means that, in an experiment to observe the transition without adding any perturbation to the pipe, the transition Reynolds number depends on background fluctuations, {\it i.e.},  it depends on the detailed setting of the experiment. Second, even with a sufficiently strong perturbation to create tiny patches of turbulence ({\it e.g.}, higher vorticity region) known as ``puffs" \cite{wygnanski1973transition}, these puffs show sudden decaying or splitting into two, whose time scales are extremely long  \cite{faisst_eckhardt_2004,PhysRevLett.96.094501,PhysRevLett.98.014501}. 
Because of this, determining the precise value at which the puffs start to sustain was for a long time an unsolvable task.

A breakthrough came after the detailed studies of puffs that revealed that the time scales of these splitting and decaying are stochastically and independently determined \cite{hof2006finite,PhysRevLett.101.214501,de_Lozar589,avila2010transient,kuik2010quantitative}. As the Reynolds number increases, the time scale of decaying (or splitting) increases (or decreases). There is thus a special Reynolds number, $\Reyc$, in which these two time scales become equal, and below this value the decaying of puffs is dominant, but above it the splitting of puffs is dominant. 
In 2011, more than a century after Reynolds's proposition, Avila {\it et al} measured $\Reyc$ by studying these two time scales of puffs~\cite{Avila192} finding a transition Reynolds number $\Reyc$ around $2040$. The obstacle of this measurement was that these time scales became extremely long when Re was close to $\Reyc$. Avila {\it et al} overcame this difficulty by preparing a long (15 m) pipe, but in their paper, they also stated that they could not observe the puff decaying and splitting within numerical simulations for $\Rey \sim \Reyc$, due to high computational costs.

The study of the turbulent-laminar transition is difficult around $\Rey_{\rm c}$, because 
the puffs are weakly unstable~\cite{doi:10.1146/annurev.fluid.39.050905.110308}, and splitting and decaying are observed as rare events.
In fact a super-exponential increase of the puff-decaying time scale has been observed as a function of the Reynolds number \cite{hof2006finite,PhysRevLett.101.214501} and its origin has been discussed using the extreme value statistics \cite{fisher1928limiting,gumbel1935valeurs,PhysRevE.81.035304,Goldenfeld2017} and directed percolation models~\cite{PhysRevE.84.035304,shih2016ecological}, but it is still unclear if this is an effective law observed only around $\Rey_{\rm c}$ or if it can be observed beyond.  The goal of this paper is to introduce a sampling method to help this situation by accelerating the measurement of the puff decaying. For the application of this method, we use a coupled map lattice model~\cite{CHATE1988409} to describe the puff dynamics that has been proposed by Barkley \cite{PhysRevE.84.016309} (below we call it Barkley model). %, whose definition is provided in Appendix~\ref{Append_Barkley}.  
However, we stress that our method can be applied to more realistic systems, including DNS of Navier-Stokes equation and experiments.

The structure of this paper is the following. We first discuss the relation between the puff-decaying time scale and a rare-event probability (the tail of an accumulative probability function)  in Section~\ref{Sec:Fluctuations_TimeScale}.
We then introduce the sampling method that uses a feedback control of the Reynolds number in Section~\ref{Sec:ReynoldsControlled}. In Section~\ref{Sec:application}, we demonstrate the application of the method to the Barkley model, and we show that the super-exponential increase of the the puff-decaying time scale is observed up to $\Rey=2500$. Within this section, we also discuss the improvement of calculation efficiency of the method (Section~\ref{SubSec_Efficiency}). In Section~\ref{Sec:Conclusion}, we conclude this paper. We note that the detailed definition of the Barkley model is provided in Appendix~\ref{Append_Barkley}.

\section{Fluctuations and puff-decaying time scale}
\label{Sec:Fluctuations_TimeScale}
\begin{figure}
\begin{center}
\includegraphics[width=0.45\textwidth]{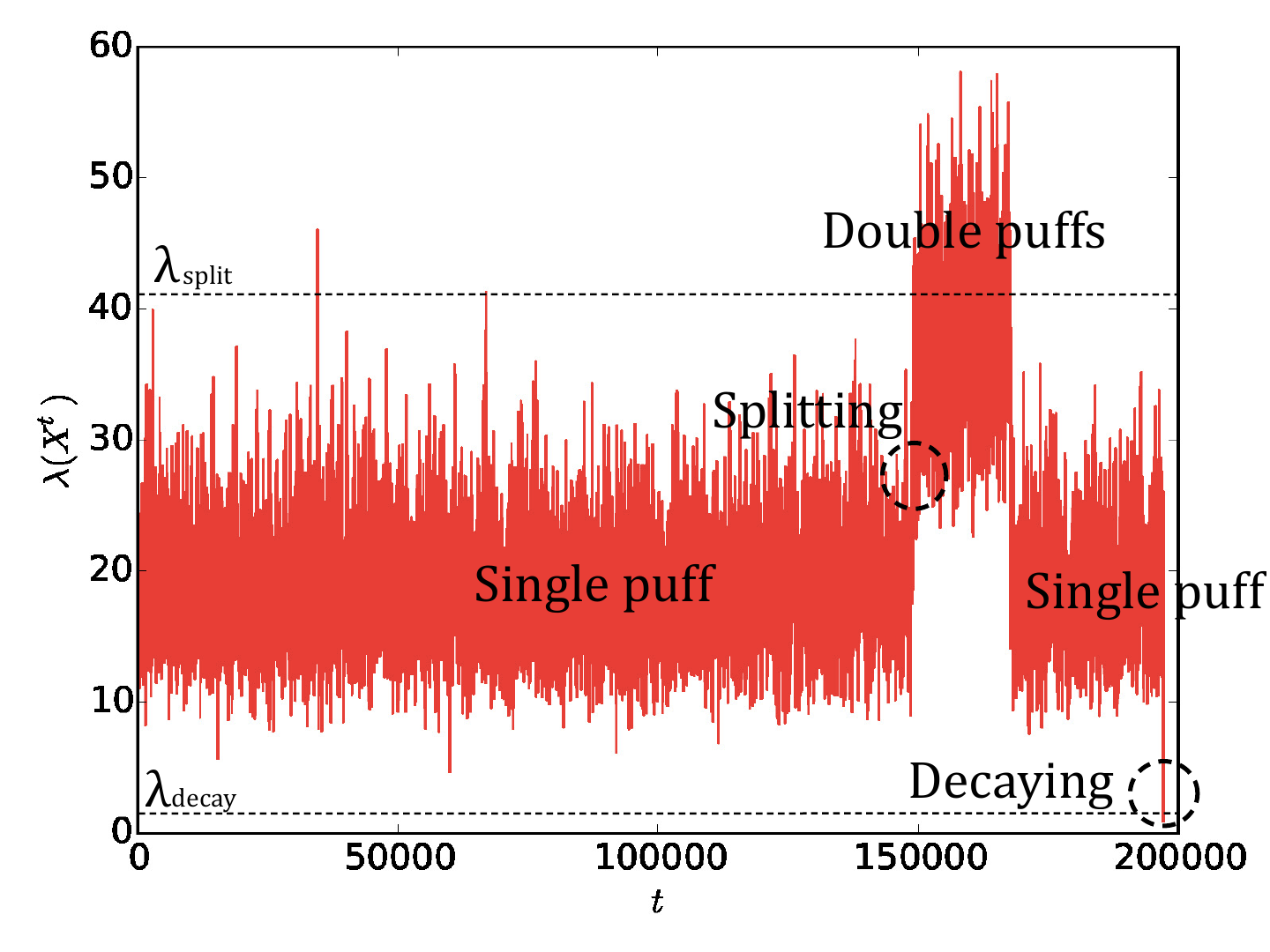} 
\caption{\label{fig:timeseries} Typical time-series data of the total turbulence intensity $\lambda(X)$ in the Barkley model \cite{PhysRevE.84.016309} with $\Rey= 2046$, showing puff splitting and puff decaying. 
When there is only a single puff, $\lambda(X)$ takes a value from 10 to 30 (approximately), but when there are two puffs, it takes a value from 30 to 60. Furthermore, as the puff decays, $\lambda(X)$ converges to 0. We thus define threshold values of $\lambda$ to judge if there exists only one puff in our pipe as $\lambda_{\rm decay} =1$ and
 $\lambda_{\rm split} = 41$, which are used throughout this paper~\cite{caption1}. Note that, although we show a decay of puff from a double-puff state to a single-puff state around $t = 1.7 \times 10^{5}$ in this figure, our measurement of puff-decaying time scale described in Section~\ref{Sec:Fluctuations_TimeScale} 
takes into account only the decay from a single-puff state.} 
\end{center}
\end{figure}

We first discuss the connection between the fluctuations of the turbulence intensity and the time scale of puff decaying.  
Let us consider a pipe flow, where we denote the velocity field of the flow
by $X$ (also by $X^t$ the field $X$ at time $t$). The total turbulence intensity is calculated from the field $X$ ({\it e.g.,} by the total energy in the radial component of $X$ or by the axial component of average vorticity), which we denote by $\lambda(X)$. For the Barkley model (whose definition is shown in Appendix~\ref{Append_Barkley}), typical dynamics of $\lambda(X)$ is presented in Fig.~\ref{fig:timeseries}. One can see that $\lambda(X)$ is fluctuating around a certain value, and $\lambda(X)$ becomes twice as big as this certain value when the puff is split into two while it takes almost 0 after the puff decays. In order to define re-laminalized (puff-decayed) states
quantitatively, we introduce a threshold value $\lambda_{\rm decay}$, such that the puff almost certainly decays once $\lambda (X)$ takes a value smaller than $\lambda_{\rm decay}$. Furthermore, in order to focus on decay events from a single-puff state, we introduce another threshold value $\lambda_{\rm split}$ that distinguishes
these two puff states  
(Fig.~\ref{fig:timeseries}).
By using these two thresholds values, we consider the following
measurement of the time-scale of puff decaying from a single-puff state.

%
% Figure dynamics
% Figure dynamics
%

(i) We start a simulation (or an experiment) to observe the turbulent puff by adding a localized perturbation to laminar flows (where only a single small puff is created). After an initial relaxation time $\tau_{\rm ini}$, we check that the puff satisfies $\lambda_{\rm decay}<\lambda(X^t)<\lambda_{\rm split}$ with $t = \tau_{\rm ini}$. We repeat (i) until we get a state that satisfies this inequality.

(ii)  During the time evolution of the puff ($t\geq \tau_{\rm ini}$), we store the value of $\lambda(X^t)$ for each time interval $\delta t_{\rm m}$. We stop this simulation when $\lambda_{\rm decay}<\lambda(X^t)<\lambda_{\rm split}$ is violated. (More precisely, we stop the simulation the first time we store $\lambda (X^t)$ after $\lambda_{\rm decay} \geq \lambda(X^t)$ or $\lambda(X^t) \leq \lambda_{\rm split}$ holds.)

(iii) When we stop the simulation, if  $\lambda_{\rm decay} \geq \lambda(X^t)$, we increment a number $n_{\rm decay}$ (that starts at 0 at the beginning of the entire measurements) by 1.  
We also increment the total number of measurements $n_{\rm tot}$ (that also starts at 0 at the beginning of the entire measurements) by $(t - \tau_{\rm ini})/\delta t_m$, where $t$ is the time when $\lambda_{\rm decay}<\lambda(X^t)<\lambda_{\rm split}$ becomes violated.

After repeating this measurement many times, we get the 
estimate of the decaying time scale $T_{\rm d}$ as 
\begin{equation}
T_{\rm d} = \frac{n_{\rm tot}  \delta t_m }{ n_{\rm decay}}.
\label{eq:Td_}
\end{equation}
In~\cite{hof2006finite,PhysRevLett.101.214501,PhysRevE.84.016309}, the puff-decaying time scale is measured from an exponential fitting 
to the probability distribution function of (each) puff-decaying time. Different from their measurements, our estimator (\ref{eq:Td_}) directly gives the expected value of the puff-decaying time. 
(Our estimator is equivalent to the one used in~\cite{hof2006finite,PhysRevLett.101.214501,PhysRevE.84.016309} when $n_{\rm decay}$ is sufficiently large.) In many experiments and numerical simulations, it has been observed that $T_{\rm d}$ 
scales in a super-exponential way as a function of $\Rey$ \cite{hof2006finite,PhysRevLett.101.214501}, {\it i.e.}, a measurement of 
$T_{\rm d}$ based on brute-force calculations becomes harder as the Reynolds number increases.

$T_{\rm d}$ is connected to rare fluctuations of the turbulence intensity. To see this, we define an accumulative probability function of $\lambda(X)$ as follows: by denoting the obtained (total) time series of $\lambda$ by $\lambda^i$ ($i = 1,2,\dots,n_{\rm tot}$), we define
\begin{equation}
P(\lambda) = \frac{1}{n_{\rm tot}}\sum_{i = 1}^{n_{\rm tot}}  
\ \theta (\lambda - \lambda^i),
\label{eq:defPlambda}
\end{equation}
where $\theta (\lambda)$ is the Heaviside step function: $\theta (\lambda) = 1$ for $\lambda > 0$ and $\theta (\lambda) = 0$ for $\lambda \leq 0$. By definition, we have $P(\lambda_{\rm decay})=n_{\rm decay}/n_{\rm tot}$. 
From (\ref{eq:Td_}), we find
\begin{equation}
P(\lambda_{\rm decay}) = \frac{\delta t_{\rm m}}{T_{\rm d}},
\label{eq:pminus}
\end{equation}
namely, the tail value of the accumulative probability $P(\lambda)$ is connected to the inverse of the puff-decaying time scale.

\begin{figure}
\begin{center}
\includegraphics[width=0.45\textwidth]{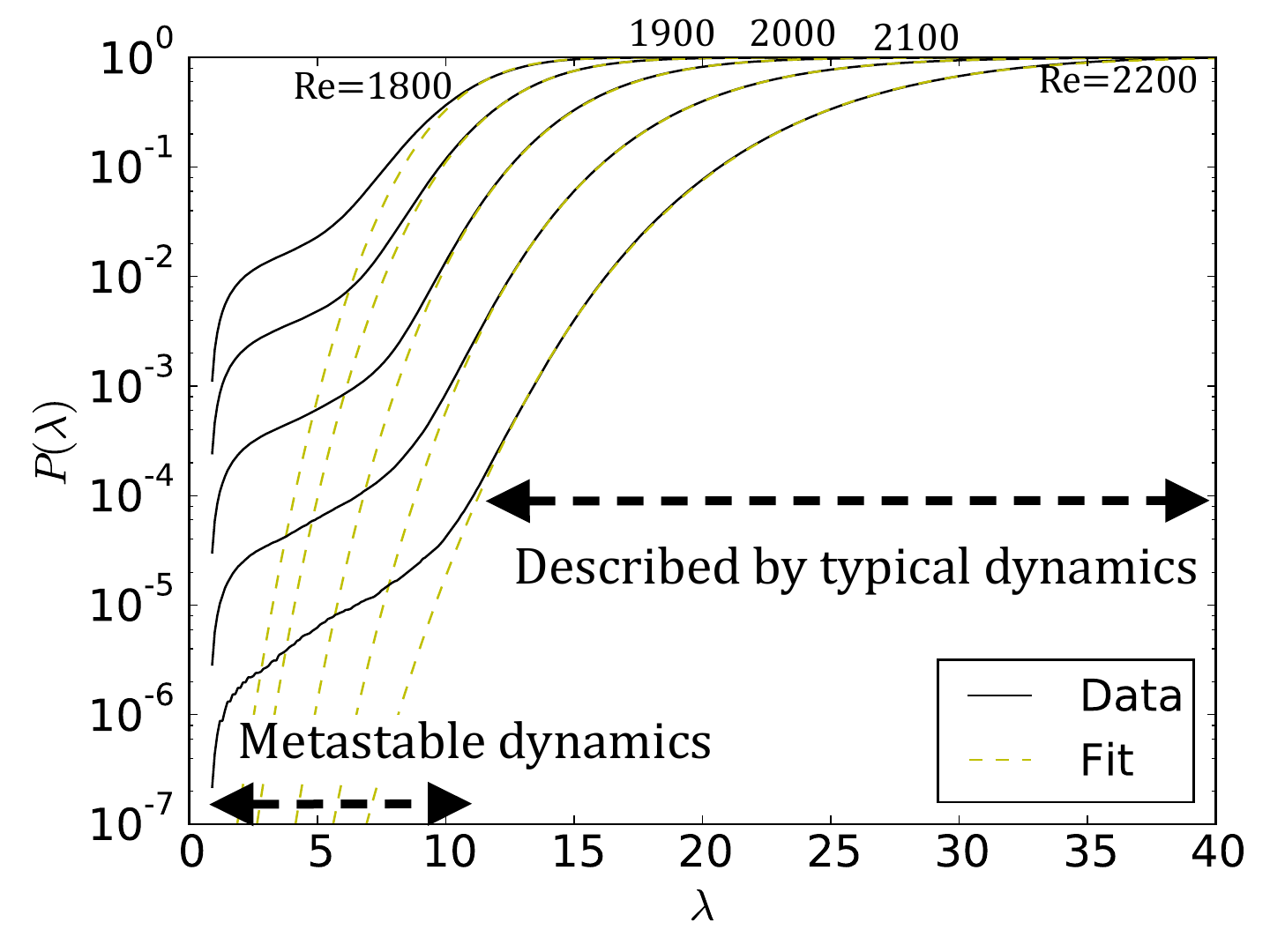} 
\caption{\label{fig:BruteForce_Q_} The accumulative probability function $P(\lambda)$ for several Reynolds numbers in the Barkley model~\cite{PhysRevE.84.016309} obtained from brute-force measurements. $P(\lambda)$ shows two different behaviors, namely (i) the one described by typical dynamics of the puffs and (ii) the one described by relatively stable dynamics (metastable dynamics) before the puffs decay. For the typical part, we fit to the data a super-exponential function defined as (\ref{eq:fitexp}), which shows good agreement with the typical part of $P(\lambda)$. } 
\end{center}
\end{figure}

\section{Reynolds number controlled procedure}
\label{Sec:ReynoldsControlled}
To measure the tail of $P(\lambda)$ efficiently, we propose a simple procedure
to control the Reynolds number during the measurement. 
In Fig.~\ref{fig:BruteForce_Q_}, we show numerical examples of $P(\lambda)$ in the Barkley model~\cite{PhysRevE.84.016309} for several Reynolds numbers. One can see that the domain of this probability function is separated into two parts: large-$\lambda$ part and small-$\lambda$ part. The large-$\lambda$ part is described by the typical dynamics, whereas the small-$\lambda$ part is described by the dynamics of atypically small puffs. 
In the small-$\lambda$ part, the slope of $P(\lambda)$ (in logarithmic scale) is smaller than the one in the large-$\lambda$ part.
This observation suggests the existence of a relatively stable state for small puffs before decaying, which we call {\it metastable state} in this paper.   
What we propose is a procedure to change the Reynolds number to efficiently create such a metastable state.

%
% Figure dynamics
\begin{figure}
\begin{center}
\includegraphics[width=0.40\textwidth]{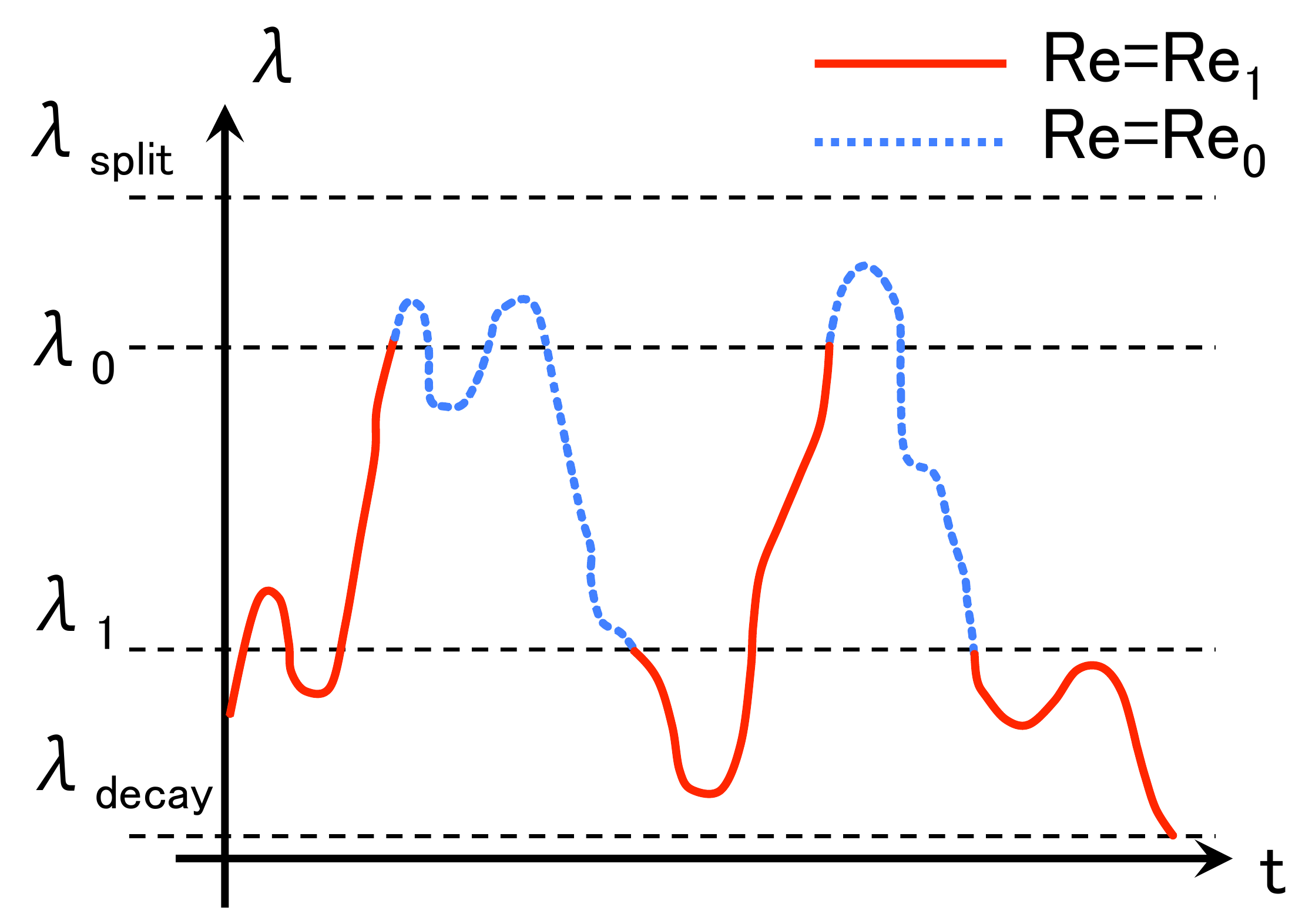} 
\caption{\label{fig:schematic} Schematic figure to explain the procedure to control the Reynolds number (Re-control) during the measurement. When $\lambda$ crosses $\lambda_1$ (or $\lambda_0$), we change the Reynolds number to $\Rey_1$ (or to $\Rey_0$), where $\Rey_1 > \Rey_0$ and $\lambda_1 < \lambda_0$. The accumulative probability of $\lambda$ in this procedure is our estimator for the tail of $P(\lambda)$.   
} 
\end{center}
\end{figure}
% Figure dynamics
%

Let us suppose that we want to study the tail of $P(\lambda)$ at $\Rey = \Rey_1$. We define another Reynolds number $\Rey_0$ that is smaller than $\Rey_1$ ($\Rey_0 < \Rey_1$), where a puff tends to become small easily. 
We also define two special values of the turbulence intensity, $\lambda_{0}$ and $\lambda_1$ ($\lambda_0 \geq \lambda_1$), at which we switch the Reynolds number. More precisely, during the procedures (i) and (ii) explained in the previous section, the following control of the Reynolds number (Re-control) is performed: 
{\it 
we set the Reynolds number to $\Rey_1$ when $\lambda(X^t)$ crosses $\lambda_{1}$ and to $\Rey_0$ when $\lambda(X^t)$ crosses $\lambda_{0}$.
} We show a schematic figure to explain this control in Fig.~\ref{fig:schematic}.  
After finishing this procedure, we collect the time-series data of $\lambda(X)$ (in the same way as the brute-force method) and calculate
the accumulative probability function of $\lambda$, which we denote by $P_{\rm tail}(\lambda)$.
What we expect is that this functional shape of $P_{\rm tail}(\lambda)$ can provide a good approximation of the correct probability $P(\lambda)$ for small $\lambda$ (tail of $P(\lambda)$). More precisely, we expect $P(\lambda)\simeq C P_{\rm tail}(\lambda)$ for $\lambda \lesssim \lambda^*$ with two constants $C$ and $\lambda^*$, which are determined by the following conditions:
\begin{equation}
CP_{\rm tail}(\lambda^*) = P(\lambda^*),
\label{eq:Pch_1}
\end{equation}
\begin{equation}
C \frac{ dP_{\rm tail}(\lambda^*)}{ d\lambda}  = \frac{dP(\lambda^*) }{ d\lambda}.
\label{eq:Pch_2}
\end{equation} 
After determining these constants, our estimator of $P(\lambda)$ is
\begin{equation}
P(\lambda) \simeq \begin{cases}
C P_{\rm tail} (\lambda)   &    {\rm for \ }  \  \lambda  < \lambda^* \\
P(\lambda) &   {\rm for \ }  \ \lambda \geq \lambda^*.
 \end{cases}
 \label{eq:PchPfit_expectation}
\end{equation}
Note that obtaining $P(\lambda)$ for $\lambda \geq \lambda^*$  is easier than obtaining the full shape of $P(\lambda)$ from brute-force calculations. Finally, we obtain the estimator of the decaying time scale $T_{\rm d}$
in our method as
\begin{equation}
T_{\rm d}\simeq \frac{\delta t_{m}}{C  P_{\rm tail}(\lambda_{\rm decay})}
 \label{eq:PchPfit_expectation2}
 \end{equation}
from (\ref{eq:pminus}).

%%%%%%%%%%%%%%%%%%%%%%
\begin{table*}
\begin{center}
          \caption{\label{Table:Parameter_setting}  Criterions to choose the parameters $\lambda_0$, $\lambda_1$ and $\Rey_0$}
          \begin{tabular}{c||c}
\vphantom{\Big|}                                             
                                   &   Condition  
\\[1mm]  \hline  \hline \vphantom{\Big|}  
Higher transition value & $\lambda_0 = \bar \lambda_{\Rey_1}$     
\\[-1mm]                       		
$\lambda_0$               & ($ \bar \lambda_{\Rey_1}$ is the average value of $\lambda$ for $\Rey_1$) 	              	                  
\\ [1mm]  \hline \vphantom{\Big|}  
Lower transition value   &  $\lambda_{\rm ms}^{\Rey_1} \ll \lambda_1 < \lambda_0 - \sqrt{2 \sigma_{\Rey_1}}$ 	 
\\[-1mm]
$\lambda_1$  	            &  ($\lambda_{\rm ms}^{\Rey_1}$ is the boundary value between the metastable and typical
\\ 
                     	            &  regions (Fig~\ref{fig:BruteForce_Q_}) for $\Rey = \Rey_1$. $\sigma^{\Rey_1}$ is a variance of $\lambda$ for $\Rey_1$) 	   	 
\\[1mm]   \hline  \vphantom{\Big|}  
Smaller Reynolds number   &	     $\Rey_0 > \Rey_1 -  \delta \Rey^*$              
\\[-1mm]
$\Rey_0$  	            &    ($\delta \Rey^*$ is a constant around $200 \sim 300$) %If $\Rey_0$ does not satisfy this  
%\\
%  	            &    %condition, the estimator of $T_{\rm d}$ gives a lower bound of real $T_{\rm d}$)	
\\[1mm] \hline
          \end{tabular}  
        \end{center}
 \end{table*}

%%%%%%%%%%%%%%%%%%%%%%
\begin{table*}
\begin{center}
          \caption{\label{Table:ParameterValues} Estimated values of $\bar \lambda_{\Rey_1}, \sqrt{2 \sigma_{\Rey_1}}$ and $\lambda_{\rm ms}^{\Rey_1}$}
          \begin{tabular}{c||c|c|c|c}
\vphantom{\Big|}                                             
                           &   $\Rey_1 = 2100$  & $\Rey_1 = 2200$ & $\Rey_1 = 2300$ & $\Rey_1 = 2400$
\\[1mm]  \hline  \hline \vphantom{\Big|}  
$\bar \lambda_{\Rey_1}$    &  21.746 ($\pm 0.003$)                & 27.5005 ($\pm 0.0005$)           & 32.4543 ($\pm 0.0001$) & 35.6467 ($\pm 0.0002$)
\\ [1mm]  \hline \vphantom{\Big|}  
$\sqrt{2 \sigma_{\Rey_1}}$ &  6.760 ($\pm 0.002$)              	 & 7.5069 ($\pm 0.0004$)          & 6.4556 ($\pm 0.0001$)  & 5.0249 ($\pm$0.0001)
\\[1mm]   \hline  \vphantom{\Big|}  
$\lambda_{\rm ms}^{\Rey_1}$ &	    $\sim 11$             & $\sim 13$      & $\sim 15$ & $\sim 20$
\\[1mm] \hline
          \end{tabular}  
        \end{center}
 \end{table*}
%%%%%%%%%%%%%%%%%%%%%%

\section{Application to Barkley model} 
\label{Sec:application}

In this section, we apply Re-control method to a model of puff dynamics proposed by Barkley \cite{PhysRevE.84.016309}. To this end, in Section~\ref{Subsec:howtochoose}, we first discuss how to choose three parameters $\lambda_0$, $\lambda_1$ and $\Rey_0$ appearing in the method. The criterion to choose them
are also summarized in Table~\ref{Table:Parameter_setting}. 
We then show the results of the application in Section~\ref{Subsec:verification}, 
followed by the discussion on how much
the method accelerates the measurement of the
time scale $T_d$  in Section~\ref{SubSec_Efficiency}. 

\subsection{Parameters $\lambda_0$, $\lambda_1$ and $\Rey_0$}
\label{Subsec:howtochoose}

\subsubsection{Criterion for $\lambda_0$}
In the method,  the Reynolds number is set to a smaller value $\Rey_0$ from the target Reynolds number $\Rey_1$ at $\lambda = \lambda_0$ in order to suppress the growth of puff and to weaken it. 
But if $\lambda_0$ is too small, the puff does not have enough time
to evolve in the target Reynolds $\Rey_1$ and is suppressed before its equilibration. We thus set the value of $\lambda_0$ to be equal or larger than the typical value of $\lambda$ in the target  Reynolds number $\Rey = \Rey_1$. More precisely, by introducing a probability density $p(\lambda)$ as
\begin{equation}
p(\lambda) = \frac{d}{d \lambda}P(\lambda),
\label{lambda_}
\end{equation}
we denote the average value of $\lambda$ for the Reynolds number $\Rey$ by
\begin{equation}
\bar \lambda_{\Rey}  = \int d  \lambda \ p(\lambda) \lambda.
\end{equation}
We then assign a condition to $\lambda_0$ as
\begin{equation}
\lambda_0 = \bar \lambda_{\Rey_1}. 
\label{eq:cond_lambda0}
\end{equation}
Note that although this condition may be weakened as $\lambda_0 \geq \bar \lambda_{\Rey_1}$, we use (\ref{eq:cond_lambda0}) for the simplicity of the argument. We stress that calculating $\bar \lambda_{\Rey_1}$ is not difficult, since it does not require the tail values of the probability $P(\lambda)$. 
Numerical examples of $\bar \lambda_{\Rey}$ are provided in Table~\ref{Table:ParameterValues}.

\subsubsection{Criterion for $\lambda_1$}

After changing the Reynolds number from $\Rey_1$ to $\Rey_0$, the puff is weakened and finally reaches a state that takes $\lambda = \lambda_1$. We then change the Reynolds number from $\Rey_0$ to $\Rey_1$. We expect that the puff quickly forgets how it 
is prepared and the statistics for $\lambda < \lambda_1$ obtained afterwords is equivalent to the 
brute-force results (in the sense of (\ref{eq:PchPfit_expectation})). For this, we discuss the lower and upper 
bounds of the parameter $\lambda_1$ as follows.

We first discuss the upper bound. When $\lambda_1$ is too large ({\it i.e.}, too close to $\lambda_0$), 
the puff often goes back to $\lambda_0$ before equilibrating. The method is not efficient
in this case, since many failed attempts are needed to get an equilibrated puff that can explore
$\lambda < \lambda_1$.  
In order to prevent this, 
we assign the upper bound of $\lambda_1$ as 
\begin{equation}
\lambda_1 < \lambda_0 - \sqrt{2 \sigma_{\Rey_1}},
\label{eq:cond_lambda1_1}
\end{equation} 
where $\sigma_{\Rey}$ is the variance of $\lambda$ calculated from the probability distribution $p(\lambda)$ as
\begin{equation}
\sigma_{\Rey} = \int d\lambda \ p(\lambda)  \left ( \lambda - \bar \lambda_{\Rey} \right )^2.   
\end{equation}
Numerical examples of $\sigma_{\Rey}$ are shown in Table~\ref{Table:ParameterValues}.

Next, we discuss the lower bound. 
If the value of $\lambda_1$ is in the metastable range of Fig.~\ref{fig:BruteForce_Q_} ({\it i.e.}, too small), 
the puff determines to decay from the configuration before equilibrated after $\Rey$ is changed to $\Rey_1$ at $\lambda = \lambda_1$. These artificial decays
carry the information of the lower Reynolds number $\Rey_{0}$ and thus bias the obtained statistics. To prevent this, 
we set the lower bound of $\lambda_1$ as
\begin{equation}
\lambda_ 1 \gg \lambda_{\rm ms}^{\Rey_1},
\label{lambda_1_lambda_MS}
\end{equation}
where $\lambda_{\rm ms}^{\Rey_1}$ is the boundary value between the metastable and typical regions of $P(\lambda)$ for $\Rey = \Rey_1$. 
Within brute-force simulations, this value is determined as the maximum value of $\lambda$ where
the super-exponential fit (which is (\ref{eq:fitexp}) in the next subsection) cannot describe $P(\lambda)$. 
Estimating such an exact value is difficult since it requires the information of the metastable part of $P(\lambda)$. Without knowing this metastable part, what we can get is the higher bound of $\lambda_{\rm ms}^{\Rey_1}$, which we denote $\tilde \lambda_{\rm ms}^{\Rey_1}$. This fact is fortunately compatible with the 
condition (\ref{lambda_1_lambda_MS}): we can get a weaker inequality using such a higher bound  by simply replacing  $\lambda_{\rm ms}^{\Rey_1}$ in (\ref{lambda_1_lambda_MS}) by  $\tilde \lambda_{\rm ms}^{\Rey_1}$, {\it i.e., } the practical condition is $ \lambda_ 1 > \tilde \lambda_{\rm ms}^{\Rey_1}$.
Rough estimations of $\lambda_{\rm ms}^{\Rey_1}$ are provided in Table~\ref{Table:ParameterValues}.

\subsubsection{Criterion for $\Rey_0$}

By choosing $\lambda_0, \lambda_1$ following the conditions (\ref{eq:cond_lambda0}), (\ref{eq:cond_lambda1_1}), and (\ref{lambda_1_lambda_MS}) above, we expect that 
(\ref{eq:PchPfit_expectation}) is satisfied if $\Rey_0$ is sufficiently close to $\Rey_1$, {\it i.e.},
\begin{equation}
\Rey_0 > \Rey_1 -  \delta \Rey^*
\end{equation} 
with a constant $\delta \Rey^*$. 
From numerical simulations for a broad range of $\Rey_1$, 
what we observe is that there indeed exists such a threshold value $\delta \Rey^*$, which is 
around $200 \sim 300$ %irrespective of the value of $\Rey_1$ 
(See Fig.~\ref{fig:Td} in Section~\ref{Subsec:decaying_time_scale} for $\Rey_0$ dependence of the estimator $T_{\rm d}$). 
To derive such a threshold value $\delta \Rey^*$ 
based on a theory seems difficult, which remains as an important open question.

%
% Figure dynamics
\begin{figure*}
\begin{center}
\includegraphics[width=0.45\textwidth]{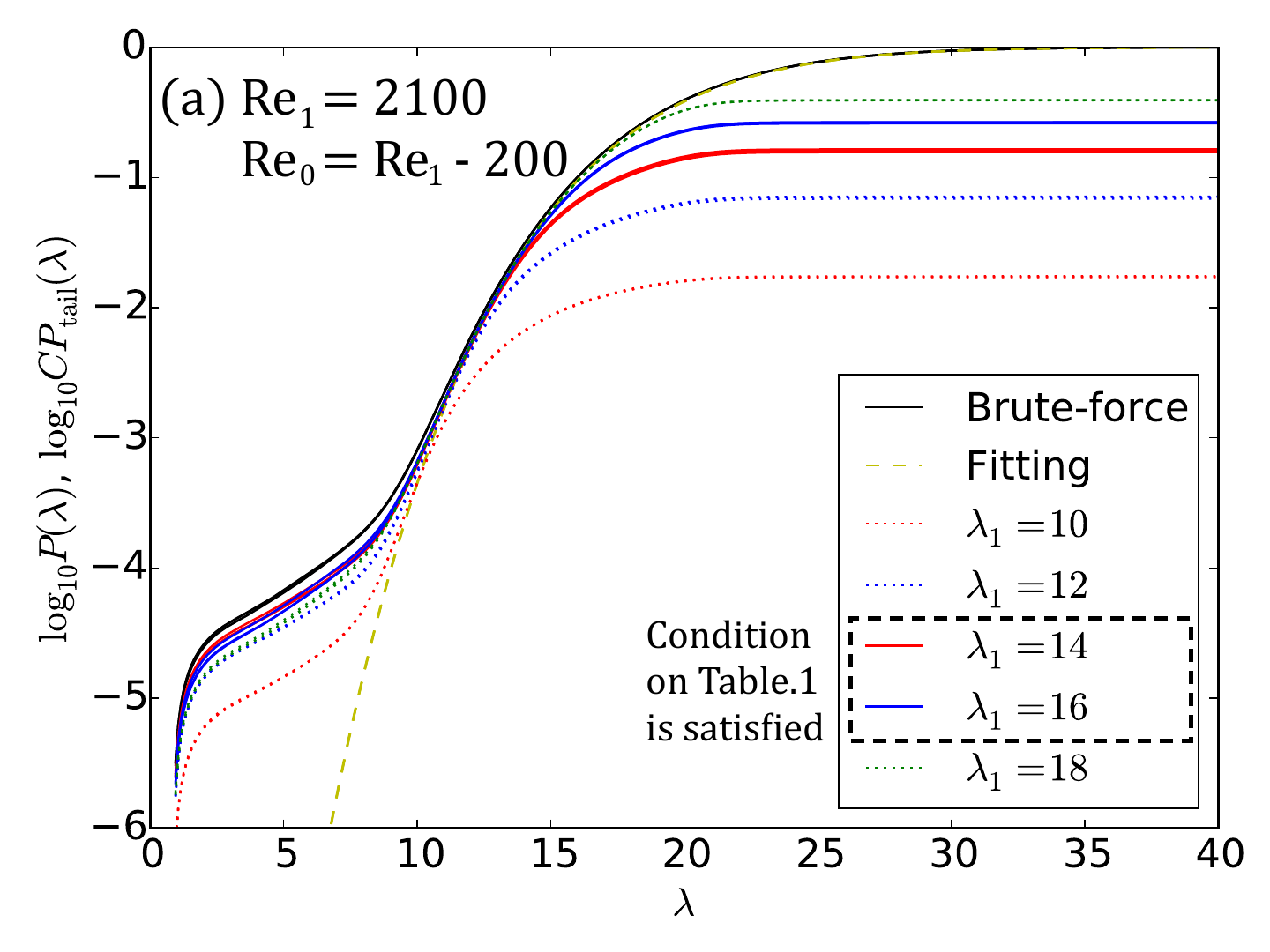} 
\includegraphics[width=0.45\textwidth]{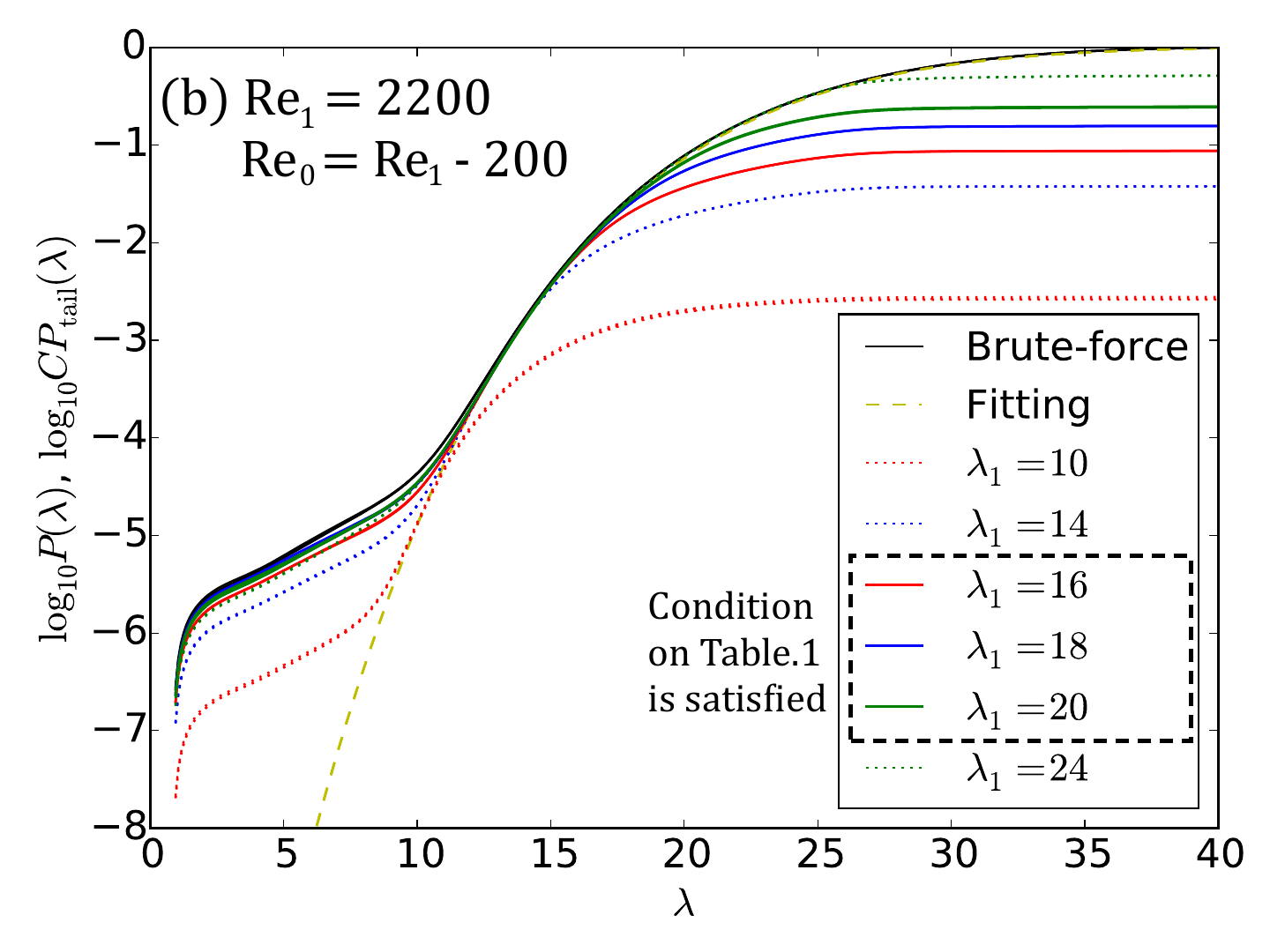} 
\includegraphics[width=0.45\textwidth]{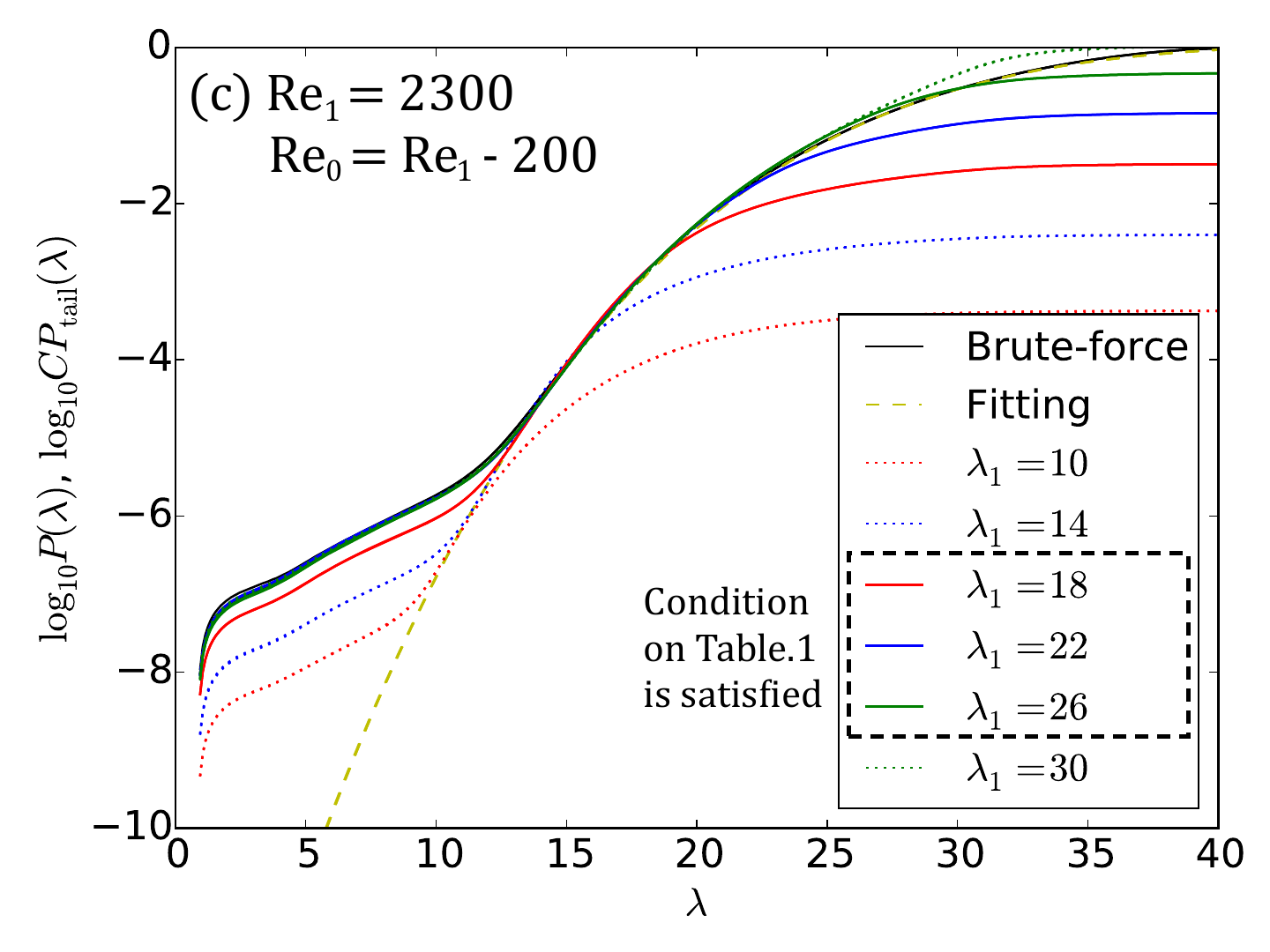} 
\includegraphics[width=0.45\textwidth]{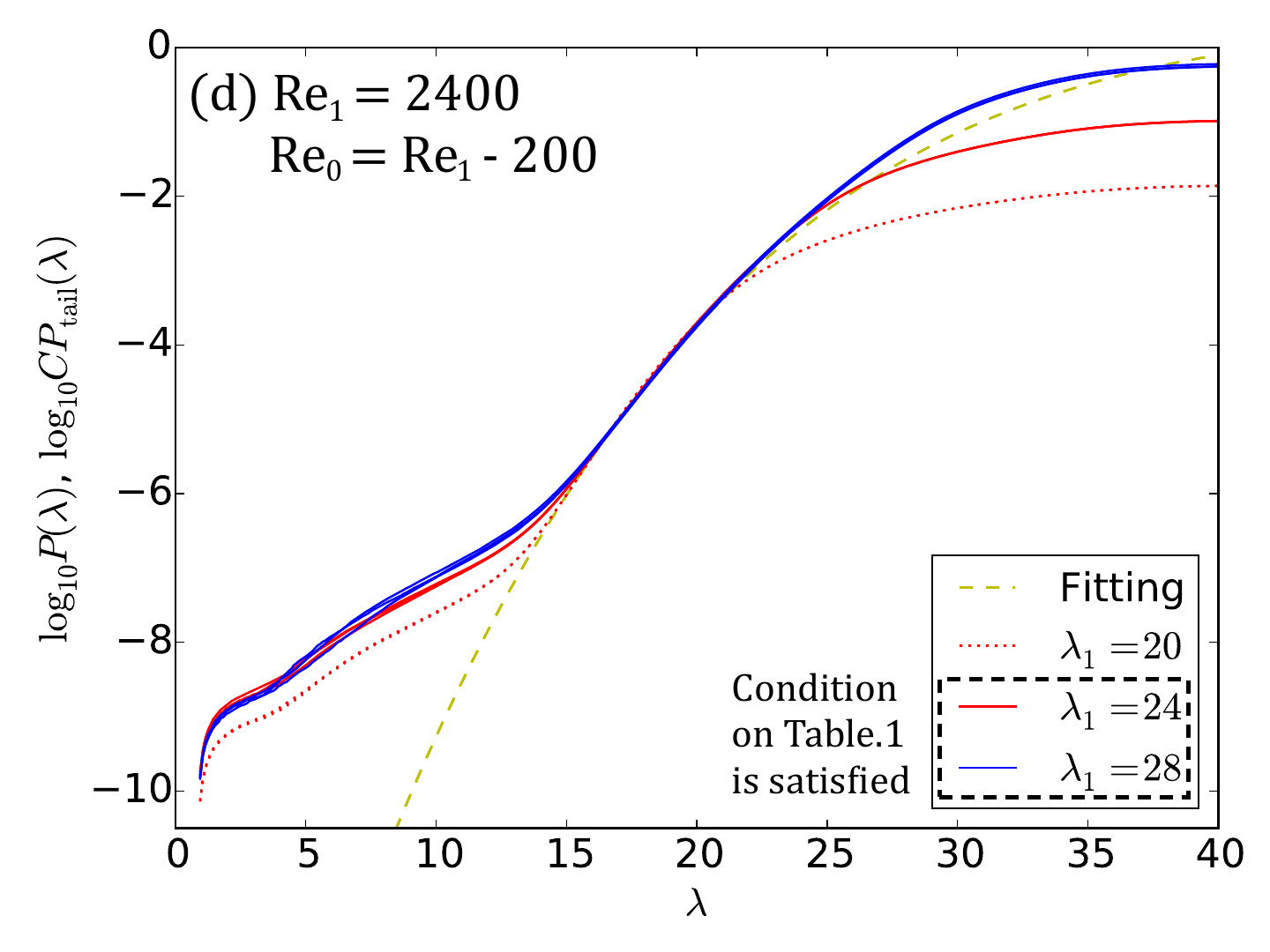} 
\caption{\label{fig:Ptail}  $\log_{10}P(\lambda)$ obtained from brute-force simulations and $\log_{10} CP_{\rm tail}(\lambda)$ obtained from Re-control method for several $\lambda_1$. 
The target Reynolds number $\Rey_1$ is set to 2100, 2200, 2300, 2400 for (a), (b), (c), (d), respectively. In each panel, we also plot the super-exponential fitting curve (\ref{eq:fitexp}). 
The parameters $\Rey_0$ and $\lambda_{0}$ are set to $\Rey_1 - 200$ and $\bar \lambda_{\Rey_1}$ according to Table~\ref{Table:Parameter_setting}, \ref{Table:ParameterValues}. Different lines in the figures correspond to different values of $\lambda_1$. 
For the panels (a-c), one can see that, within the range of $\lambda_1$ that satisfies the condition of Table~\ref{Table:Parameter_setting},
$P(\lambda)$ agrees with $CP_{\rm tail}(\lambda)$ for $\lambda < \lambda^*$ (where $\lambda^*$ is the connecting point between $\log_{10}P(\lambda)$ and $\log_{10} CP_{\rm tail}(\lambda)$). 
This demonstrates the relation  (\ref{eq:PchPfit_expectation}). For each simulation, we repeat the procedure (i-iii) in Section~\ref{Sec:Fluctuations_TimeScale} until $n_{\rm decay}$ becomes $3600$, except for 
some lines in the panels (c) and (d): in these cases, because of limited simulation time, we stop the procedure (i-iii) before $n_{\rm decay}$ reaches this value. The values of $n_{\rm decay}$ to stop the procedures are summarized
in Table~\ref{Table:Parametersused_ndecay} in Appendix~\ref{parameters_used_ndecay}. 
The statistical errors of each line are small. In order to show this, we divide the obtained data (for each line) into three sets and plot the averaged results over each set in the same figure. Three independent-realization lines are hardly distinguishable, demonstrating small statistical errors. In the panel (d), we only plotted the lines obtained from Re-control method, since the brute-force results are not converged in the tail. (``Fitting'' describes the typical part of this un-shown brute-force line). With the aid of our Re-control method, the full shape of $P(\lambda)$ can be obtained even in this case, whose tail $CP_{\rm tail}(\lambda_{\rm decay})$ corresponds to the inverse of the puff-decaying time scale (as (\ref{eq:PchPfit_expectation2})). 
} 
\end{center}
\end{figure*}
% Figure dynamics
%

\subsection{Numerical demonstration of (\ref{eq:PchPfit_expectation}):
\\  equivalence between $P(\lambda)$ and $C P_{\rm tail}(\lambda)$}
\label{Subsec:verification}

We numerically demonstrate (\ref{eq:PchPfit_expectation}). 
In order to determine the constant $C$ from the two conditions (\ref{eq:Pch_1}) and (\ref{eq:Pch_2}),
we use the shape of the typical part of $P(\lambda)$. 
In order to make sure that we do not use 
the information of the tail of $P(\lambda)$ (because it is our goal), we use the following function $P_{\rm fit}(\lambda)$ instead of $P(\lambda)$ that describes only the typical part:
\begin{equation}
P_{\rm fit}(\lambda) \propto \int_{0}^{\lambda} dx \exp \left [-\tilde \gamma (x - \tilde \lambda) - e^{-\tilde \beta(x - \tilde \lambda)} \right ],
\label{eq:fitexp}
\end{equation}
where $\tilde \gamma$, $\tilde \lambda$, $\tilde \beta$ are parameters determined by fitting to $P(\lambda)$.
(This fitting can be done without knowing the tail of $P(\lambda)$.) Examples of this function for several Reynolds numbers are shown in Fig.~\ref{fig:BruteForce_Q_}.
Note that the derivative of $P_{\rm fit}(\lambda)$ has a simpler form, which is studied in Appendix~\ref{Append_NumericalStudy}. 
To determine the constant $C$, we first fix $\lambda^*$ from the following condition 
\begin{equation}
\frac{\partial}{\partial \lambda}\log P_{\rm tail}(\lambda^*) = \frac{\partial}{\partial \lambda}\log P_{\rm fit}(\lambda^*). 
\label{eq:condition2_}
\end{equation}
More technically, we determine $\lambda^*$ that minimizes (LHS - RHS)$^2$ of (\ref{eq:condition2_}). After determining $\lambda^*$, we then calculate $C$ from
\begin{equation}
C=\frac{P_{\rm fit}(\lambda^*)}{P_{\rm tail}(\lambda^*)}. 
\end{equation}
It is straightforward to see if these $C$ and $\lambda^*$ satisfy (\ref{eq:Pch_1}) and (\ref{eq:Pch_2}).

We plot $CP_{\rm tail}(\lambda)$ obtained in this way in Fig.~\ref{fig:Ptail}
for several target Reynolds numbers: $\Rey_1 = 2100, 2200, 2300, 2400$. We choose the parameters $\lambda_0, \Rey_0$
following the criterion discussed in the previous subsection (summarized in Table~\ref{Table:Parameter_setting} together with Table~\ref{Table:ParameterValues}) for several $\lambda_1$. We also plot $P(\lambda)$ obtained from brute-force simulations in the same figure. 
One can see that $CP_{\rm tail}(\lambda)$ agrees with $P(\lambda)$ for $\lambda < \lambda^*$
when $\lambda_1$ satisfies the criterion.

%
% Figure dynamics
\begin{figure}
\begin{center}
\includegraphics[width=0.45\textwidth]{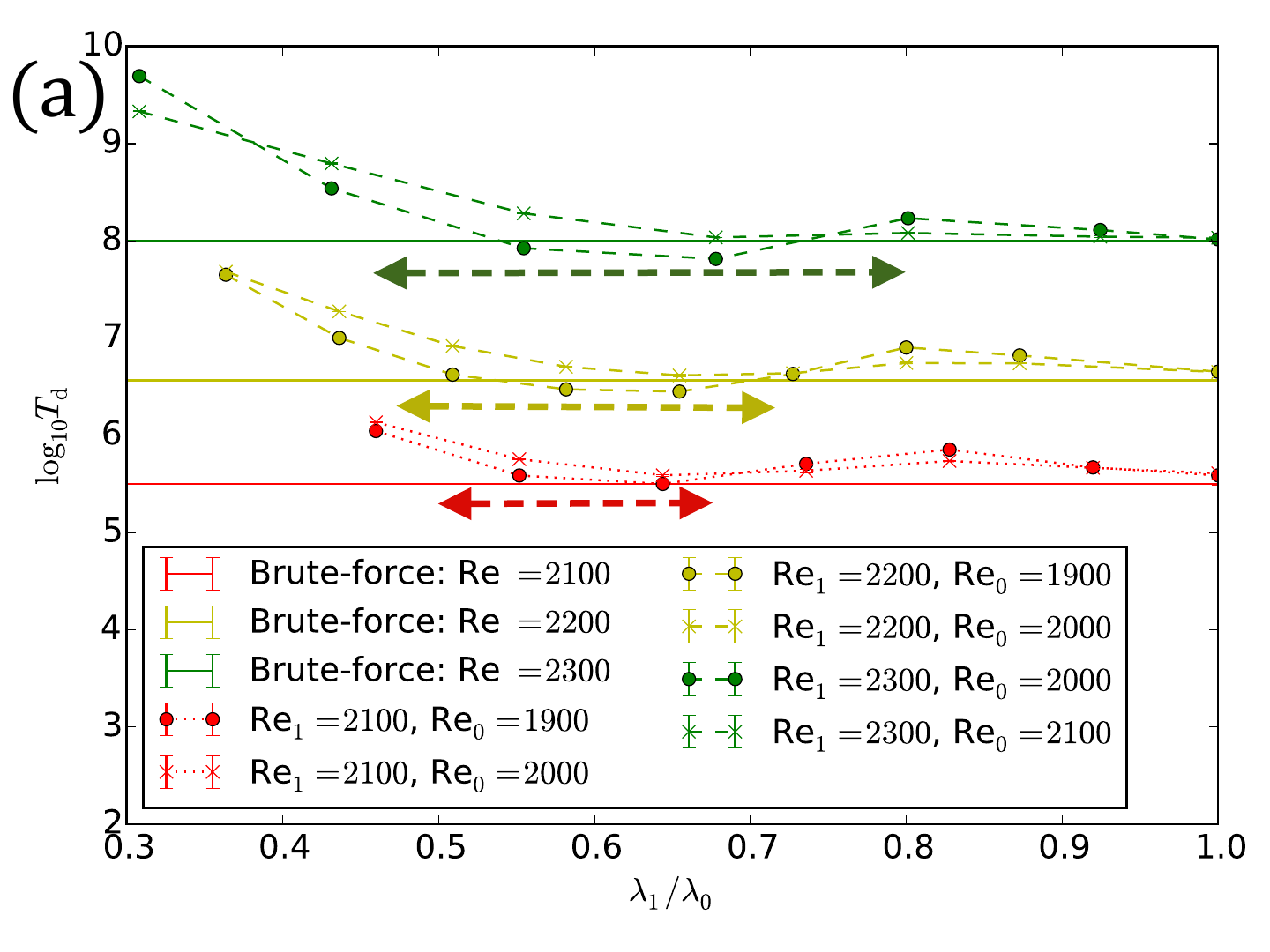} 
\includegraphics[width=0.45\textwidth]{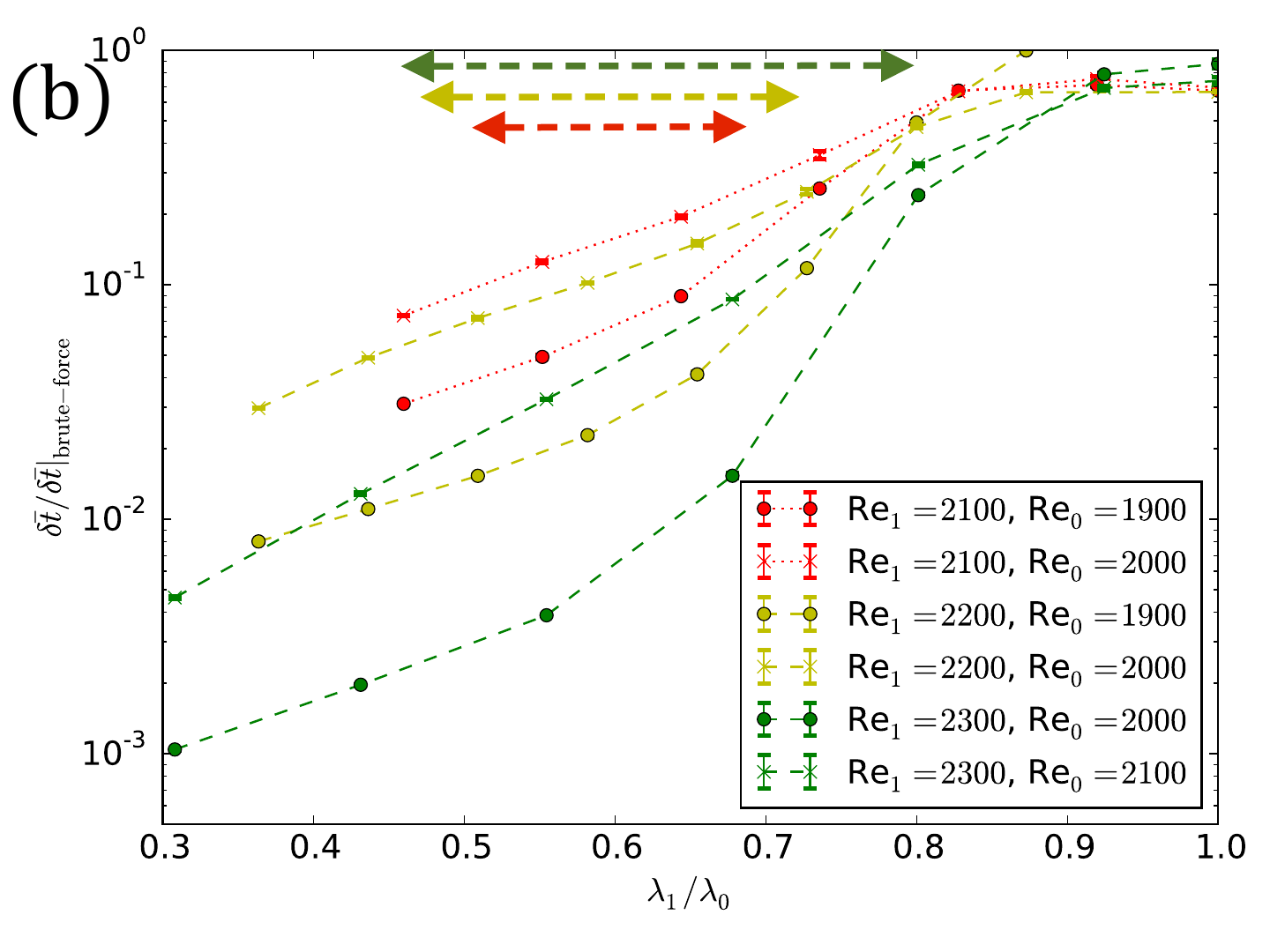} 
\caption{\label{fig:Td} {\bf (a)} The logarithm of puff-decaying time scale $\log_{10} T_{\rm d}$ obtained from Re-control method as a function of the parameter $\lambda_1/\lambda_0$, (where $\lambda_0$ is fixed to $\bar \lambda_{\Rey_1}$, whose value is given in 
Table~\ref{Table:ParameterValues}). The brute-force estimation of $T_{\rm d}$ is also shown in the same figure as 
solid lines. By using dashed double-headed arrows, we indicate the range of $\lambda_1/\lambda_0$ in which
the condition of Table~\ref{Table:Parameter_setting} is satisfied. 
In this range, one can see that the estimators of $T_{\rm d}$ in brute-force and Re-control methods agree well. 
{\bf (b)} The average simulation time $\delta \bar t$ to observe one decaying event for Re-control methods, divided by the same quantity for the brute-force method $\delta \bar t|_{\rm brute-force}$. How much faster is Re-control method than the brute-force one is given as the inverse of this quantity. In the range of $\lambda_1$ where the condition of Table~\ref{Table:Parameter_setting} is satisfied, this value takes less than 1, meaning that Re-control method is more efficient than the brute-force method.  
}  
\end{center}
\end{figure}
% Figure dynamics
%

\subsection{Puff decaying time scale}
\label{Subsec:decaying_time_scale}

In Fig.~\ref{fig:Td}(a), we plot the puff-decaying time scale $T_{\rm d}$ as a function of $\lambda_1/\lambda_0$, obtained from $C P_{\rm tail}(\lambda_{\rm decay})$ by using (\ref{eq:PchPfit_expectation2}) (where we set $\delta t_m = 1$). 
We also plot $T_{\rm d}$ obtained from brute-force simulations. One can see that the estimator of Re-control method
agrees with the brute-force result in the range of parameters that satisfy the condition in Table~\ref{Table:Parameter_setting}. 
We note that  our estimator tends to predict larger values than the correct one if the value of $\lambda_1/\lambda_0$ is smaller than this range.

\begin{figure}
\begin{center}
\includegraphics[width=0.45\textwidth]{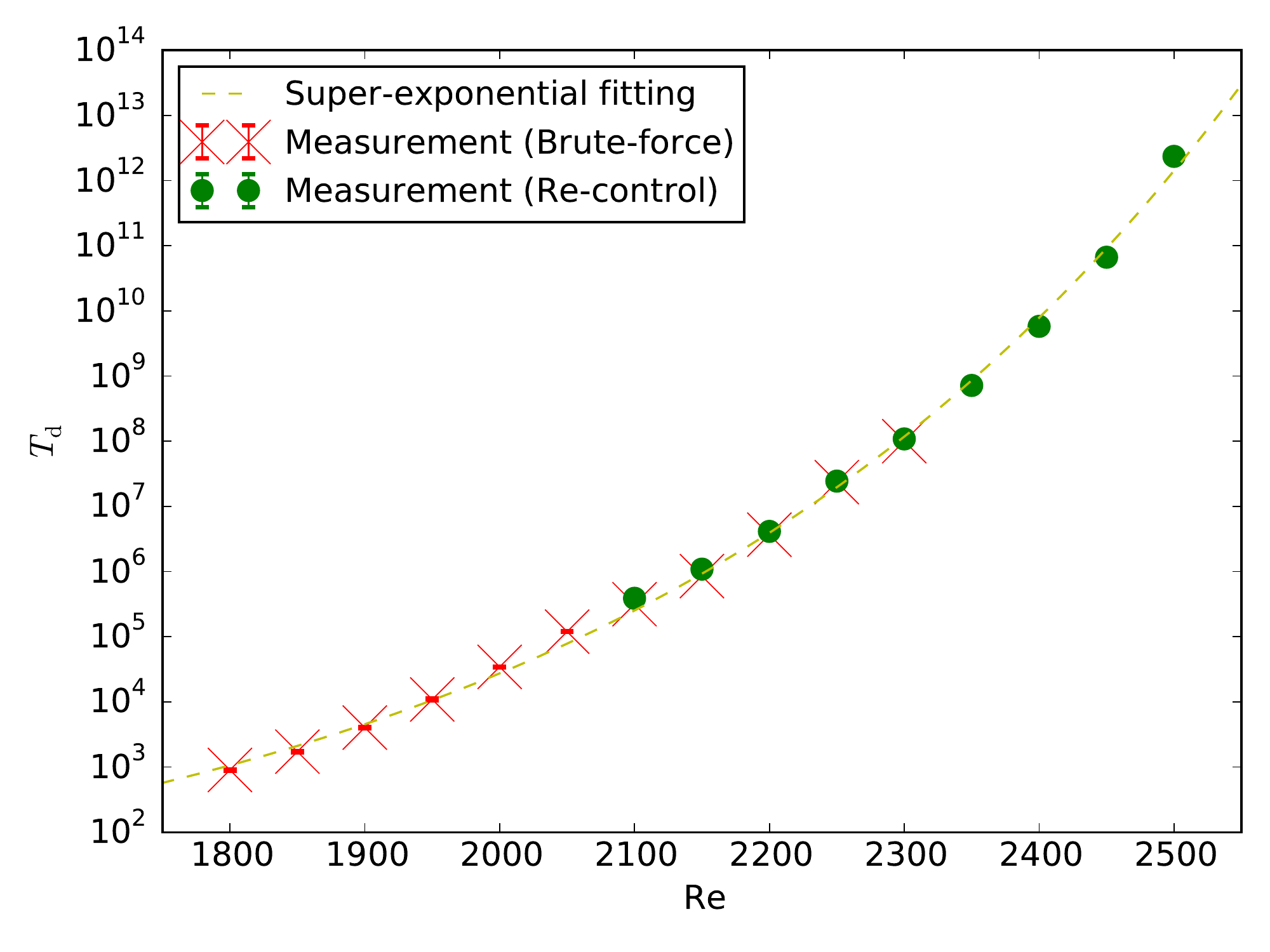} 
\caption{\label{fig:Timescale_log_large} The puff-decaying time scale $T_d$ obtained from brute-force measurements (from $\Rey=1800$ to $2300$ indicated as red crosses) and by Re-control method (from $\Rey=2100$ to $2500$ indicated as green circles). 
We stop the measurement procedures (i-iii) in Section~\ref{Sec:Fluctuations_TimeScale} when $n_{\rm decay}$ becomes 3600 for lower Reynolds numbers and much smaller values for higher Reynolds numbers. See Table~\ref{Table:Parametersused_ndecay} in Appendix~\ref{parameters_used_ndecay} for more detail.
By dividing the obtained data for each point into three sets, we estimate error bars. These error bars in the figure show small statistical errors. The data points by brute-force measurements (from $\Rey=1800$ to $2300$) and the ones by Re-control method (from $\Rey = 2350$ to 2500) are fitted by a super-exponential function defined as $\exp \left \{ \exp [ a (\Rey - b) + c \right ] \} $ with fitting parameters $a,b,c$. 
These parameters are determined using the Levenberg-Marquardt algorithm, which are $a = 2.12 \times 10^{-3}$, $b = 945$ and $c=0.82$. The obtained super-exponential function is plotted as a yellow dashed line in the figure, showing a good agreement with the data points. For Re-control method, we use the parameters $\lambda_0, \lambda_{1}, \Rey_0$ that satisfy the condition in Table~\ref{Table:Parameter_setting}. For more precise values, see Table~\ref{Table:Parametersused} in Appendix~\ref{parameters_used}. } 
\end{center}
\end{figure}

In Fig.~\ref{fig:Timescale_log_large}, we plot the puff-decaying time scale $T_d$ as a function of $\Rey$. 
The results of brute-force and Re-control methods are agree with each other for a broad range of $\Rey$. We then fit a super-exponential function to these data and plot it in the same figure.
One can see that the super-exponential curve describes well the obtained numerical data, 
supporting the existence of super-exponential law even for high Reynolds numbers. 
We expect that the small deviation of data from this super-exponential curve at $\Rey=2500$ is an 
artifact: possible reasons of this deviation are 
too small value of $n_{\rm decay}$ (Table~\ref{Table:Parametersused_ndecay} in Appendix~\ref{parameters_used_ndecay})
or $\lambda_1$ (the description in Appendix~\ref{parameters_used}), because of our  
limited simulation time.

%
% Figure dynamics
% Figure dynamics
%

\subsection{Efficiency of Re-control method}
\label{SubSec_Efficiency}

Here, we discuss how much Re-control method accelerates the measurement of the puff-decaying
time scale $T_{\rm d}$. For this, we consider 
the time duration of an entire simulation to observe one puff-decaying event in average. 
This time duration includes the preparation of  initial conditions in the procedure (i) (Section~\ref{Sec:Fluctuations_TimeScale}). We count the total time steps during the repetition 
of the procedure (i-iii), which we denote by $T_{\rm all}$. Then, the average time 
duration $\delta \bar t$ per unit decaying event is defined as
\begin{equation}
\delta \bar t = \frac{T_{\rm all}}{n_{\rm decay}}.
\end{equation}
As this number becomes smaller, one can observe more decaying events in a fixed simulation time, {\it i.e.}, obtain more statistics to evaluate the time scale of decaying events. 
We also define the same quantity for brute-force calculations, which we denote by $\delta \bar t_{\rm brute-force}$. 
In Fig.~\ref{fig:Td}(b), we plot the ratio between these two time durations: $\delta \bar t/\delta \bar t_{\rm brute-force}$. 
One can see that in the range of $\lambda_1$ that satisfies
the condition of Table~\ref{Table:Parameter_setting}, $\delta \bar t/\delta \bar t_{\rm brute-force}$ takes a value from (roughly) 0.005 to 0.5. 
Since the inverse of $\delta \bar t/\delta \bar t_{\rm brute-force}$
is the speed-up due to the method, we find that Re-control method is 2 to 200 times 
more efficient than the brute-force method. 
Note that the efficiency of the method increases as $\lambda_1$ decreases (or $\Rey_1 - \Rey_0$ increases). 
This tendency continues even if the condition in Table~\ref{Table:Parameter_setting} is not satisfied, 
although, in this case, the systematic errors from the correct result become non-negligible.

\section{Conclusion}
\label{Sec:Conclusion}
In this paper, in order to measure the puff-decaying time-scale efficiently, 
we introduce a simple procedure where the Reynolds number is controlled during the measurement.  The method does not include any complicated procedure: only changing the Reynolds number is required. We thus expect that it can be applied to DNS of Navier-Stokes equation and even to experiments.

The method is applied to the chaotic Barkley model~\cite{PhysRevE.84.016309}, and shows that the super-exponential law of the puff-decaying time scale is satisfied even for high Reynolds numbers until $\Rey = 2500$, where the puff-decaying time scale is around $10^{12} \sim 10^{13}$ and brute-force calculations cannot be used to estimate it. 
As a byproduct of the application, we find that the bulk part of $P(\lambda)$ is well-described by a super-exponential function (see Fig.~\ref{fig:BruteForce_Q_} and Appendix~\ref{Append_NumericalStudy}). 
Although this fitting function is not necessary for the application of our method, it will be interesting to see if this property holds for even more realistic systems, since the super-exponential behavior of a probability function may be the origin of the super-exponential time scale of the puff decay~\cite{PhysRevE.81.035304}.

\begin{acknowledgements}
T. N. thanks Dwight Barkley, Bruno Eckhardt, Nigel Goldenfeld, Jorge Kurchan, Bj\"orn Hof, Masayuki Ohzeki, Joran Rolland, Ohad Shpielberg and Kazumasa Takeuchi for fruitful discussions and comments. T. N. also appreciates the summer school, Fundamental Problems in Statistical Physics XIV, that provided him an occasion for several useful discussions. 
This work was granted access to the HPC resources of MesoPSL financed by the Region Ile de France and the project Equip@Meso (reference ANR-10-EQPX-29-01) of the program Investissements d'Avenir supervised by the Agence
Nationale pour la Recherche.
\end{acknowledgements}

\appendix

\section{Barkley Model}
\label{Append_Barkley}

Here we introduce a coupled map lattice model proposed by Barkley \cite{PhysRevE.84.016309} to describe the puff dynamics in pipe flows. This one-dimensional deterministic model consists of only a few hundreds of degrees of freedom, but in spite of the simplified nature of the model, it captures the basic property of puff dynamics, splitting, decaying and also the super-exponential law of the puff-decaying and -splitting
time scale.

\subsection{Definition of the model}

We consider a pipe flow modeled as follows~\cite{PhysRevE.84.016309}. We denote by $x=1,2,\dots,L$ the axial position of the pipe, and we define, at each position $x$, the axial velocity of the flows $u_x$ and the turbulence intensity (such as the axial component of the vorticity) $q_x$. These variables depend on time, which we assume discrete $t \ (= 0,1,2,\dots)$, {\it i.e., }  $u^t = (u_x^t)_{x=0}^{L}$  and $q^t =(q^t_x)_{x=0}^{L}$ for $t = 0,1,2,\dots$. We impose periodic boundary conditions to these fields: $u^t_{L+1} =u^t_1$  and $q^t_{L+1} =q^t_1$. For simplicity, we denote by $X$ the set of these two fields: $X=(q,u)$.  
We set the downstream advection speed to be 1 without loss of generality, which means that $q_{x+1}^{t+1}$ and $u_{x+1}^{t+1}$ are determined from the fields one step before at the position $x$, $q_{x}^{t}, u_{x}^t$, and their derivatives (such as $q_{x}^{t} - q_{x-1}^{t}, u_{x}^t - u_{x-1}^t, q_{x-1}^t - 2 q_{x}^t + q_{x+1}^t, \dots$).
In laminar flows, the axial velocity field $u_x^t$ takes the largest value 1 (the downstream advection speed) at all the position $x$. But in the presence of turbulence, $u_x^t$ becomes inhomogeneous, taking a value less than 1. We take into account this fact in the time evolution equation of $u_x^t$ by constructing a simple combination of these fields as follows
\begin{equation}
u_{x+1}^{t+1} = u_{x}^t + \epsilon_1 (1 - u_x^t) - \epsilon_2 u_x^t q_x^t - c (u_x^t - u_{x-1}^t),
\label{eq:evolu_u}
\end{equation}
where $\epsilon_1, \epsilon_2, c$ ($\epsilon_1 > 0, \epsilon_2 > 0, c > 0$) are parameters. The second term of this right-hand side enhances the relaminarization of flows, since this second term takes only a positive value that makes $u_{x+1}^{t+1}$ be closer to the downstream advection speed, while the third term reduces the value of $u_{x+1}^{t+1}$ due to the presence of the turbulence (non-zero value of $q_x^t$).  
The fourth term enhances the uniformity of the field $u_x^t$. When 
$u_x^t - u_{x-1}^t$ is positive (or negative), it decreases (or increases) $u_{x+1}^{t+1}$ to reduce $u_{x+1}^{t+1} - u_{x}^{t+1}$ in the next time step.

\begin{figure}
\begin{center}
\includegraphics[width=0.5\textwidth]{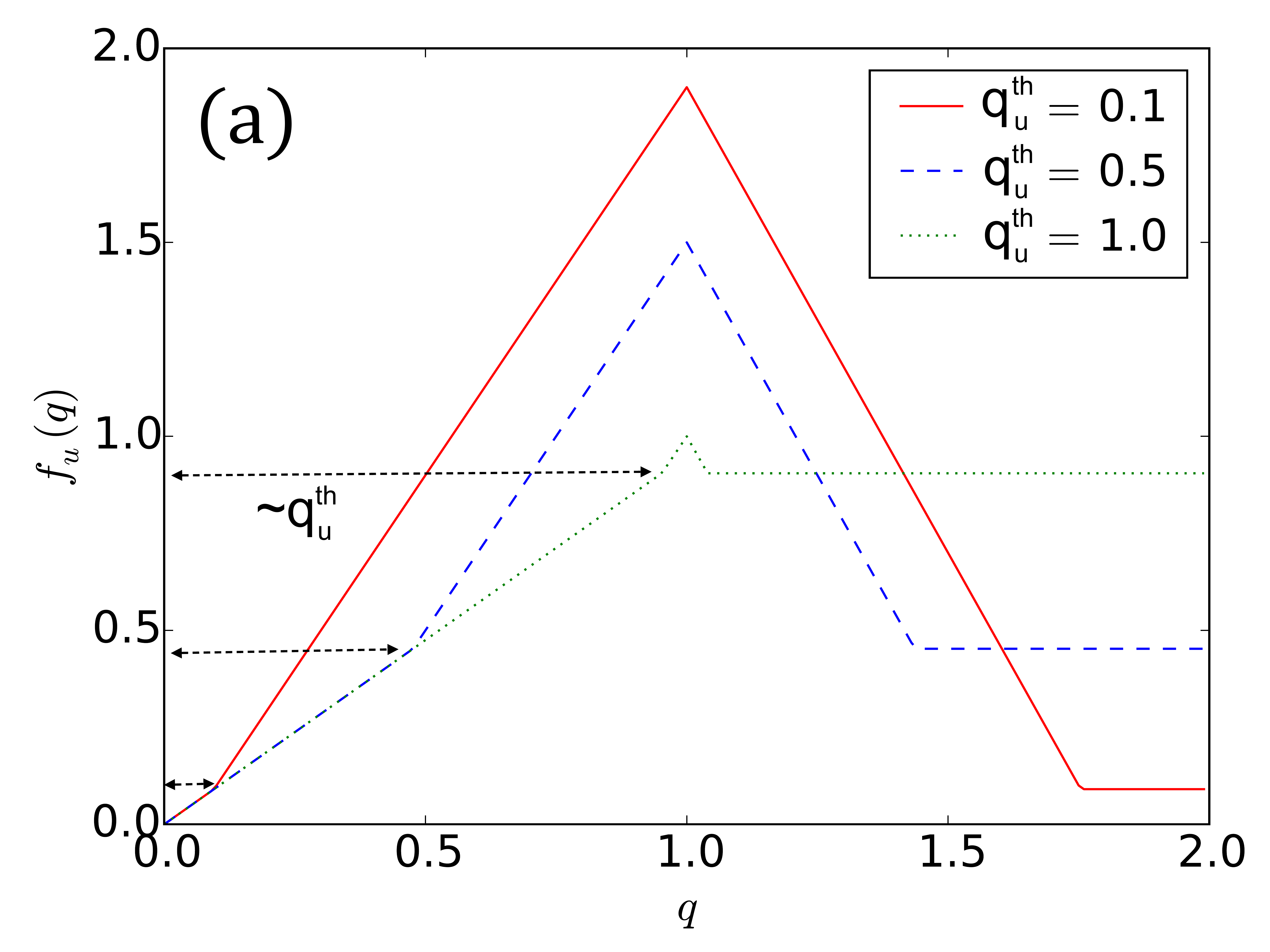}
\includegraphics[width=0.235\textwidth]{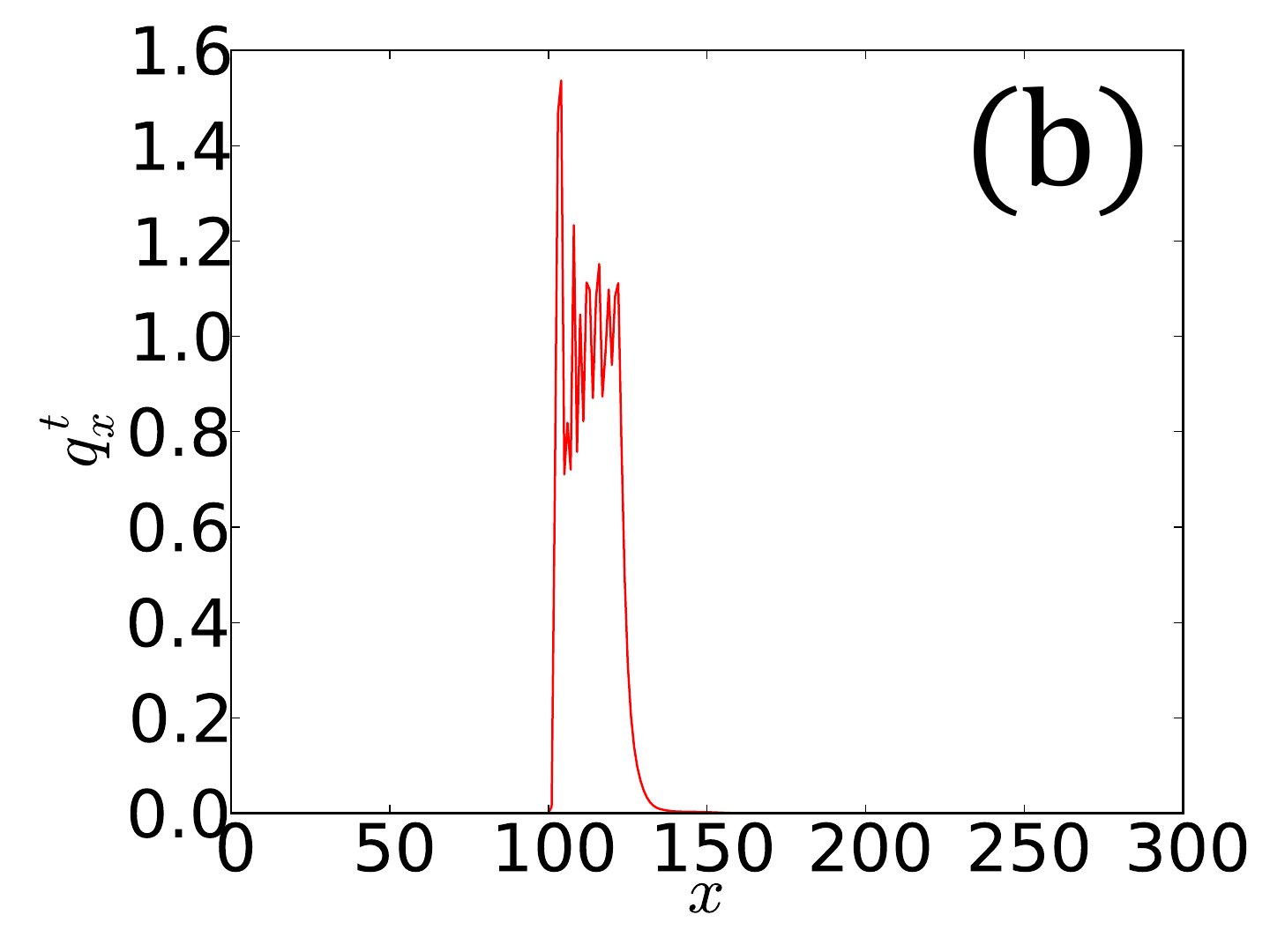} \
\includegraphics[width=0.235\textwidth]{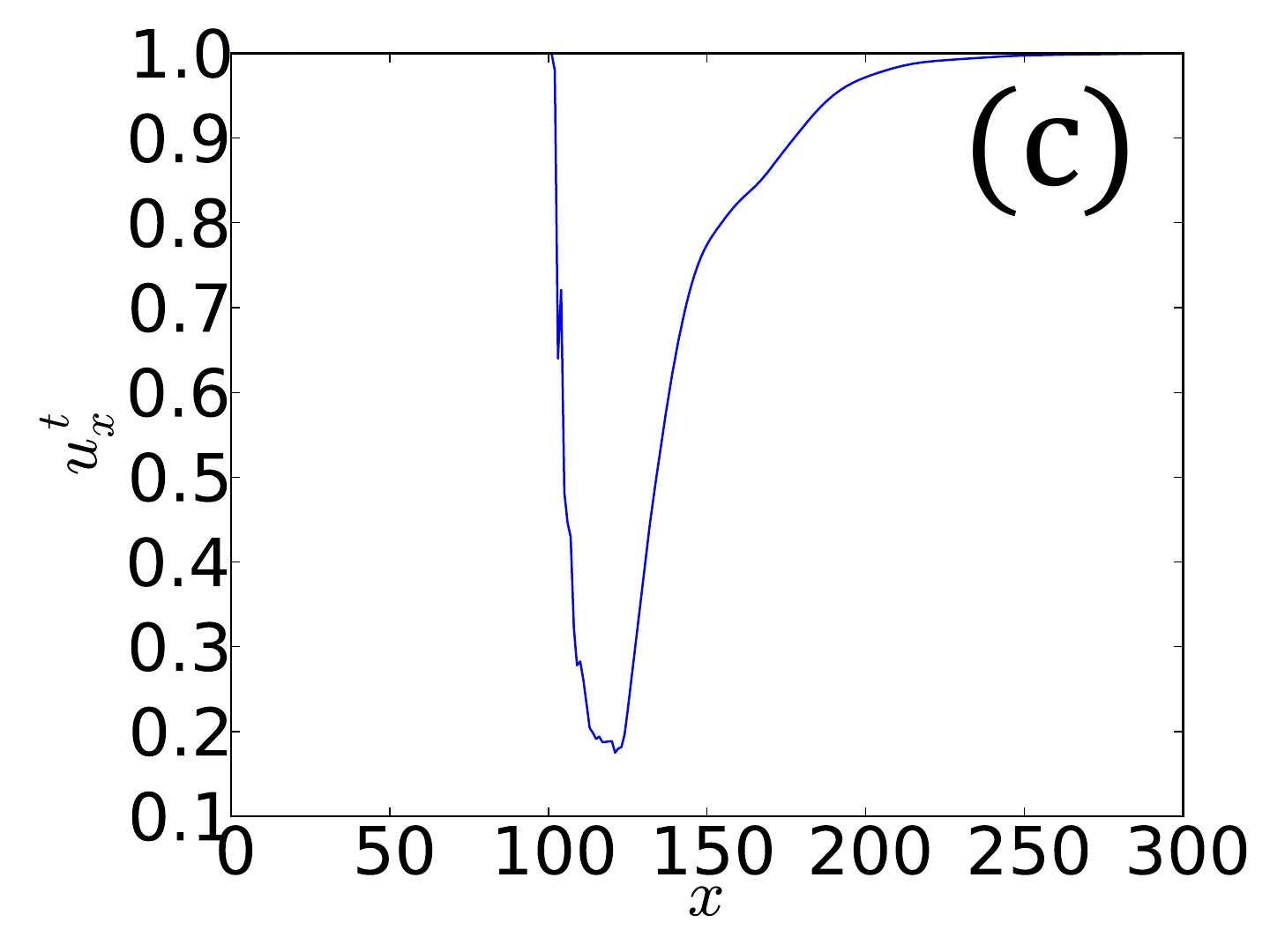} \caption{\label{fig:f_explanation} 
{\bf (a)} Functional shape of the tent map $f_u(q)$ for different values of $q^{\rm th}_u$.   
As $q^{\rm th}_u$ decreases (or increases), the size of the triangle increases (or decreases), which makes the system to be more (or less) chaotic. 
We set the Reynolds number $\Rey$ to 2046, and the rest of the parameters $d,\epsilon_1,\epsilon_2,c,\gamma,\beta$ to $0.15,0.04,0.2,0.45,0.95,0.4$ according to Ref.~\cite{PhysRevE.84.016309}.
{\bf (b,c) }Snapshots of typical configuration of $q_x^t$ (b) and $u_x^t$ (c).  } 
\end{center}
\end{figure}

For the turbulence intensity $q_x^t$, from the observation that the pipe flow turbulence is locally a chaotic repeller \cite{doi:10.1146/annurev.fluid.39.050905.110308}, we consider two types of dynamics for $q_{x+1}^{t+1}$, which are decaying dynamics and chaotic dynamics. When the turbulence intensity is locally smaller than a certain value, the time-evolution equation for the turbulence intensity in that region is a simple diffusion-like equation that enhances relaminarization. But when it is locally larger than the certain value, the time evolution is described by a chaotic map, introducing a non-trivial nature to this model.  Such a threshold value should be a function of $u_x^t$. When $u_x^t$ is large (or small), such a threshold value should be small (or large), because large (or small) axial currents easily (or hardly) induce turbulence. As the simplest manner, we define this threshold value $q^{\rm th}_u$ as a linear function of $u$ as
\begin{equation}
q^{\rm th}_{u} \equiv \frac{2000}{2-\gamma} (1 - 0.8  \ u) \Rey ^{-1},
\end{equation}
where $\gamma$ is a parameter that takes a value close to 1 (but less than 1), $\Rey$ is a parameter corresponding to the Reynolds number and $u$ is the local axial velocity, such as $u_x^t$. The constant $2000/(2-\gamma)$ is merely to adjust the scale of $\Rey$ to make the transition happen around $2040$.  
By using this threshold value, $q_{x+1}^{t+1}$ is determined as \cite{PhysRevE.84.016309}
\begin{equation}
q_{x+1}^{t+1} = F_{u_x^t}\left [  q_{x}^{t} + d (q_{x-1}^t - 2 q_{x}^t + q_{x+1}^t)  \right ],
\label{eq:evolu_q}
\end{equation}
where $d$ is a small parameter and $F_{u}[ \cdot ]$  is defined from the following map
$f_u$ as $F_{u}[ \cdot ]\equiv f_u(f_u(\cdot)))$: 
\begin{equation}
f_u(q) = \gamma q,		
\end{equation}
for $q< q^{\rm th}_{u}$ (decaying dynamics) and
\begin{equation}
f_u(q) = \begin{cases}
2 q - q^{\rm th}_{u} (2-\gamma)		&  {\rm if} \   q^{\rm th}_{u} \leq q < 1, 	\\
4 + \beta - q^{\rm th}_{u} (2-\gamma) - (2+\beta) q	 & {\rm if} \ 1 \leq q< Q_0,		\\
\gamma q^{\rm th}_{u}  & {\rm if} \  Q_0 \leq q,
\label{eq:chaoticmap}
\end{cases}
\end{equation}
for $q\geq q^{\rm th}_{u}$ (chaotic dynamics) with a constant $Q_0$ ($\equiv (4 + \beta - q^{\rm th}_{u} (2-\gamma) - \gamma Q_1)/(2+\beta)$) and a parameter $\beta$. We note that the chaotic dynamics (\ref{eq:chaoticmap}) is nothing but a tent map. To provide an insight into the map $f_u$, we show an example of $f_u$ in Fig.~\ref{fig:f_explanation}(a), 
where one can see that as $q^{\rm th}_{u}$ becomes larger, the triangle part (the tent shape part in the figure) becomes smaller, making the system to be less chaotic. 
When $q_{x}^{t} < q_u^{\rm th}$ (or more precisely $ q_{x}^{t} + d (q_{x-1}^t - 2 q_{x}^t + q_{x+1}^t)  < q_u^{\rm th}$), the time evolution equation is simply written as $q_{x+1}^{t+1} = \gamma^2 \left [  q_{x}^{t} + d (q_{x-1}^t - 2 q_{x}^t + q_{x+1}^t)  \right ]$. Since we set $\gamma < 1$, one can see that $q_x$ is diffusing with decreasing its intensity by $\gamma^2$. We note that, when all $q_x$ ($x=1,2,\dots,L$) follow such dynamics, they converge to 0.

\subsection{Numerical example}

\begin{figure*}
\begin{center}
\includegraphics[width=0.45\textwidth]{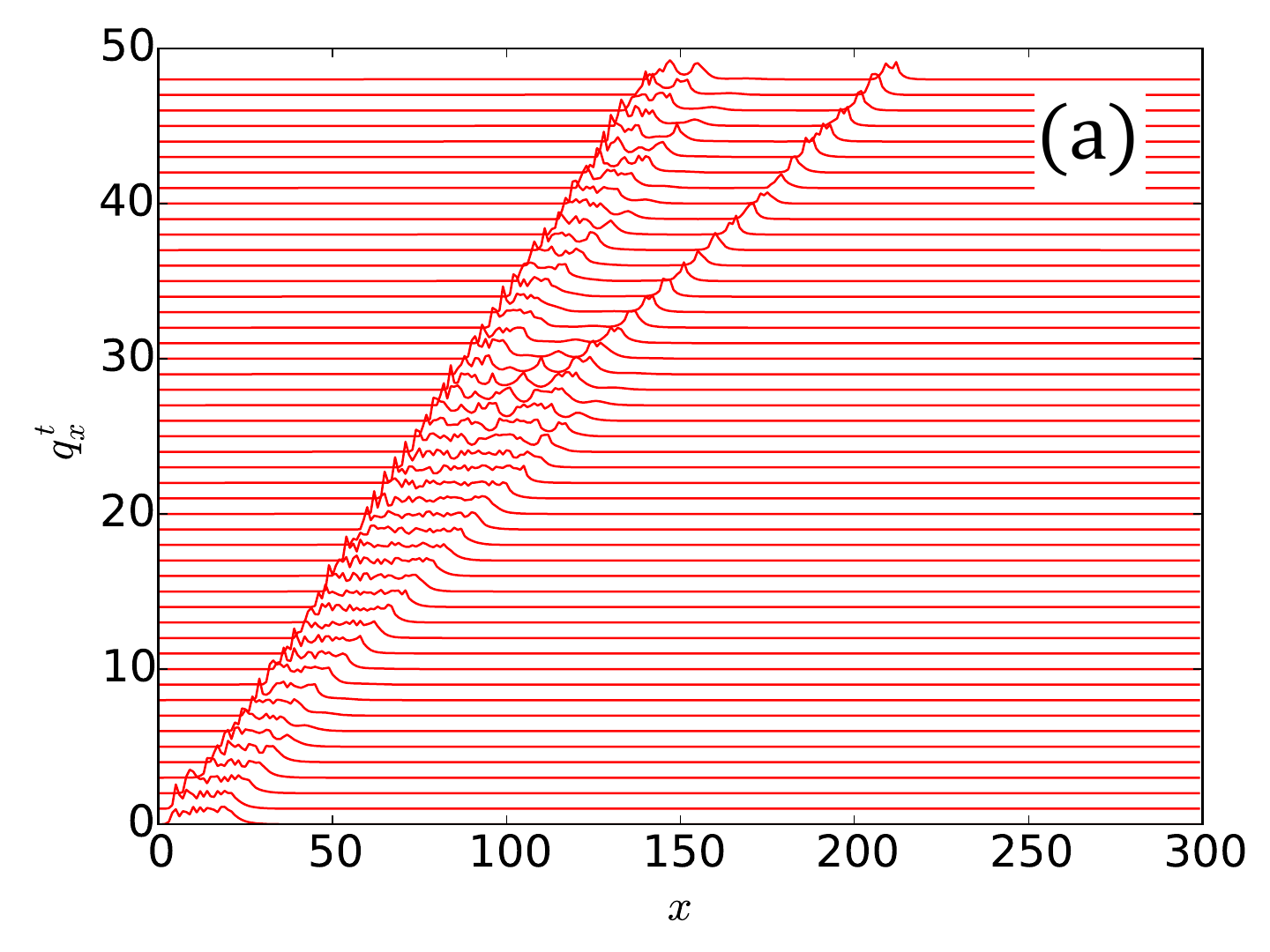} 
\includegraphics[width=0.45\textwidth]{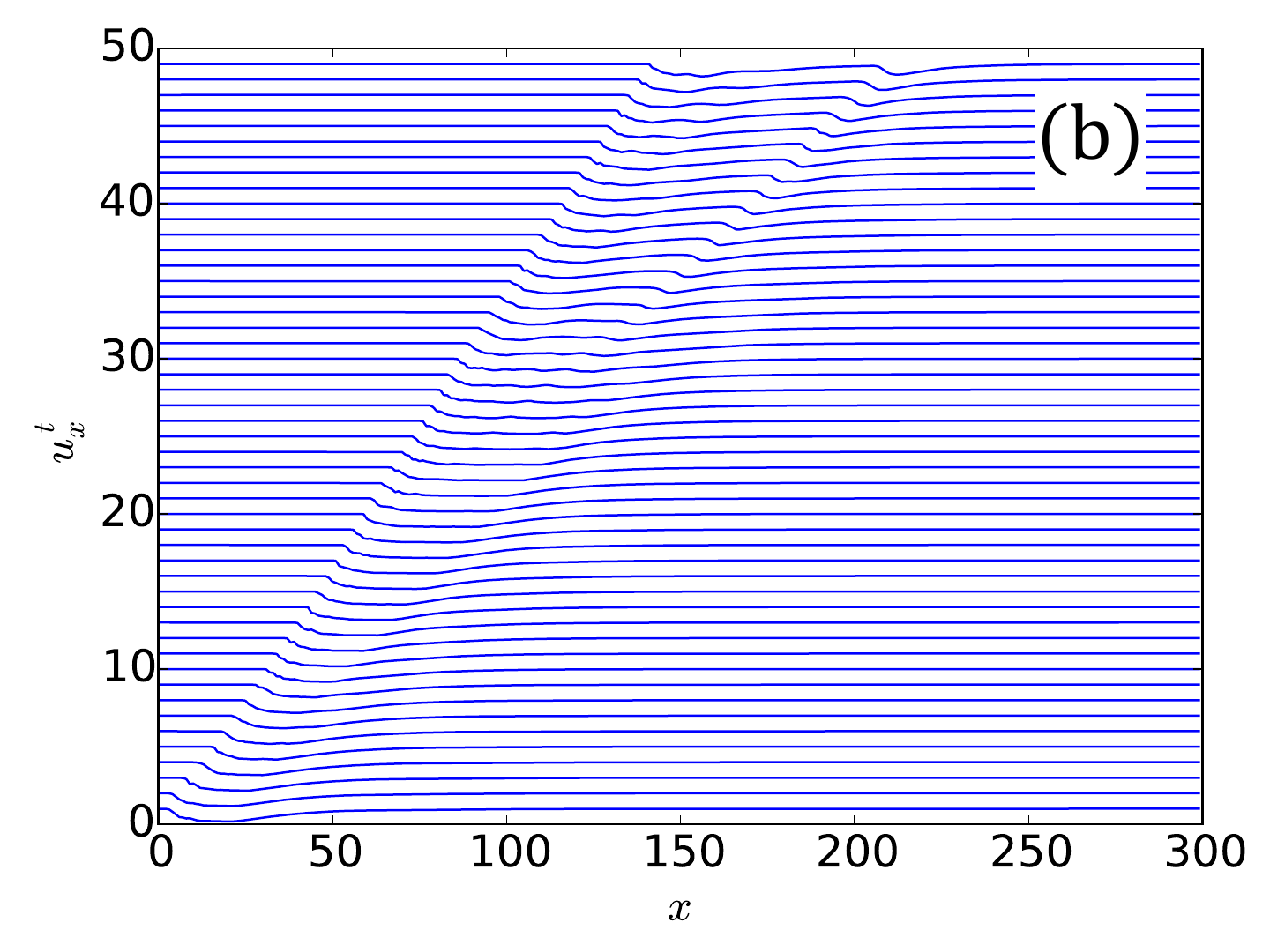} \
\includegraphics[width=0.45\textwidth]{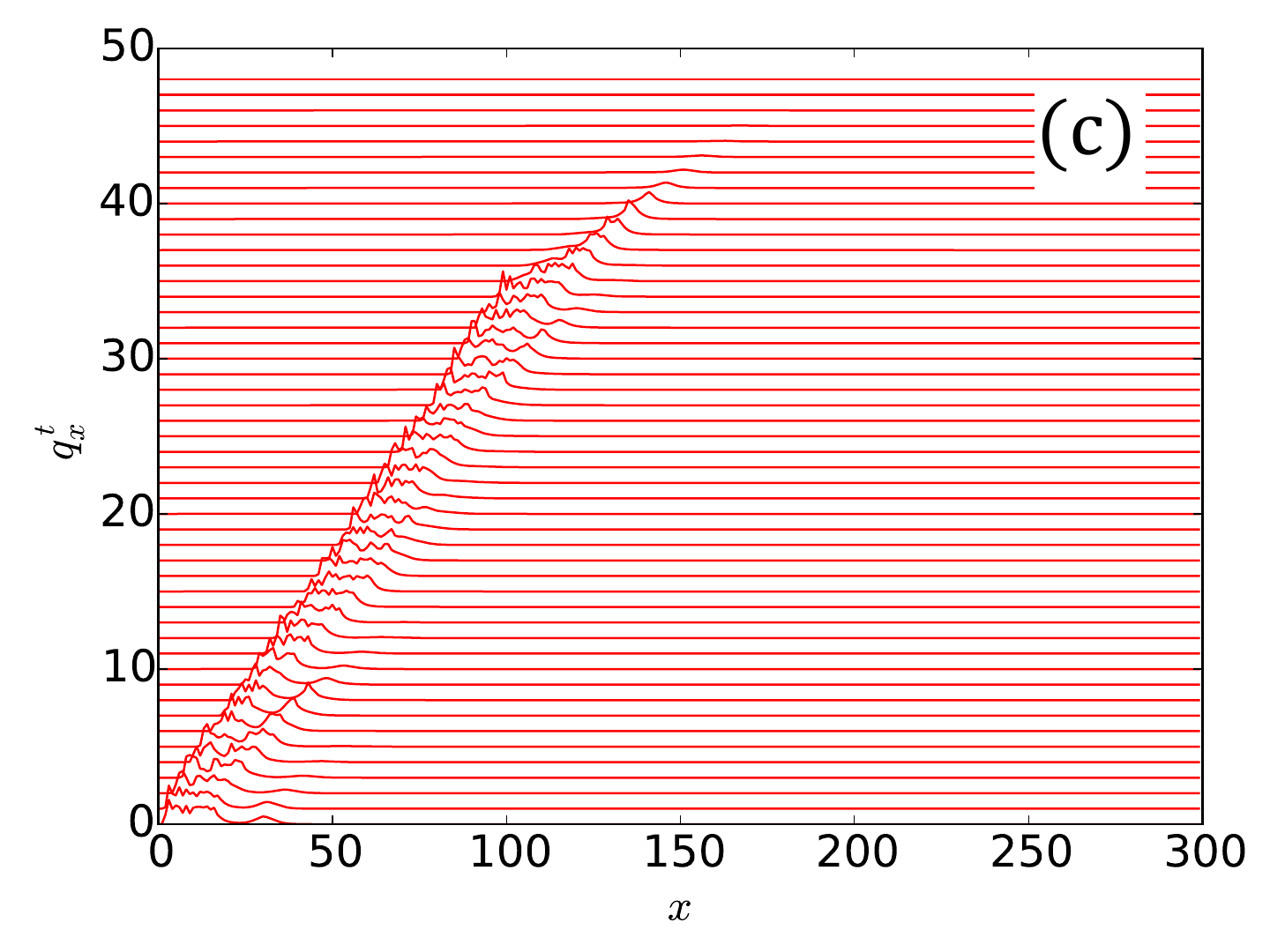} \
\includegraphics[width=0.45\textwidth]{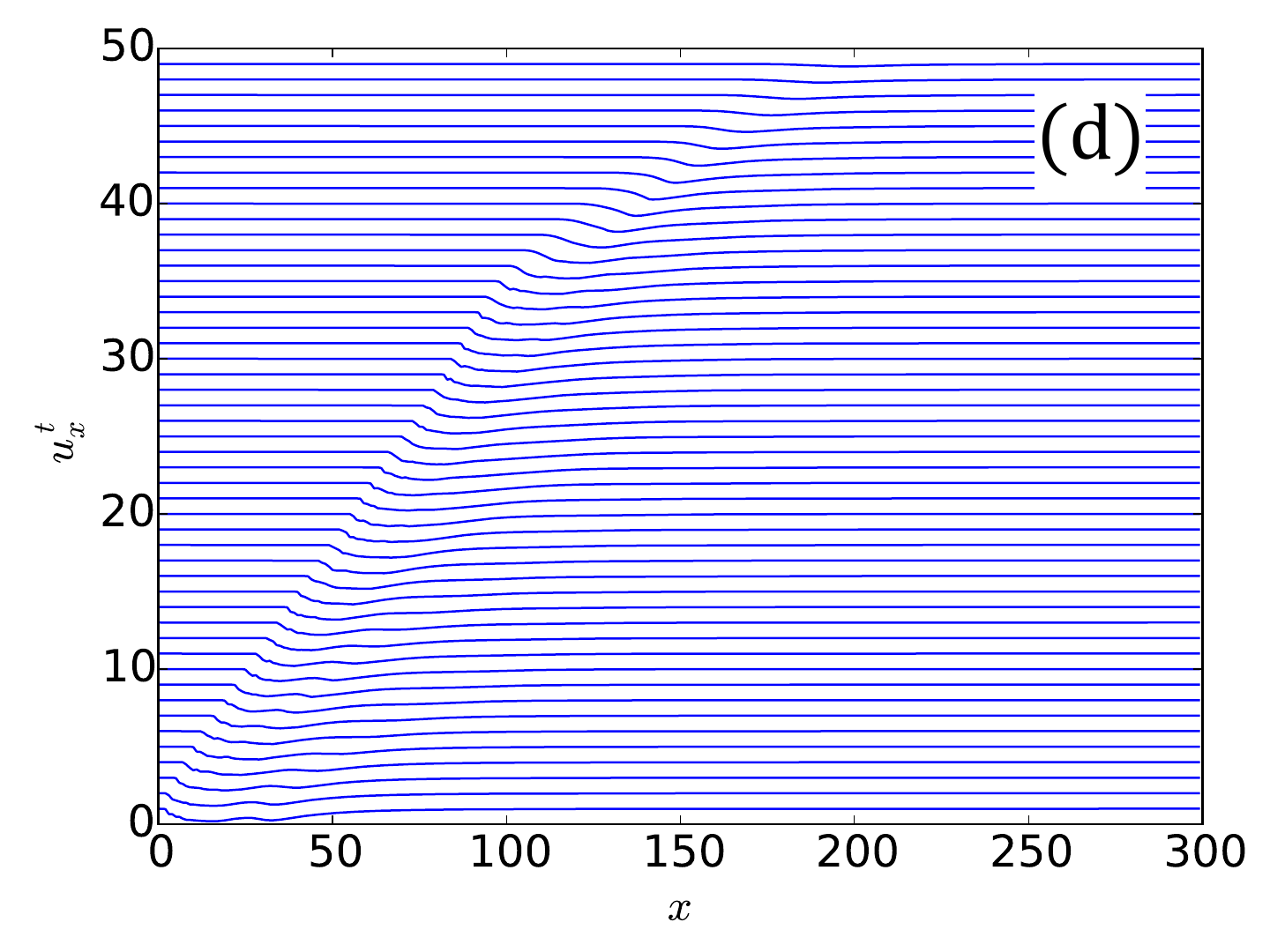} 
\caption{\label{fig:f_split} 
Snapshots of configurations $q$ [(a) and (c)] and $u$ [(b) and (d)], demonstrating puff splitting [(a) and (b)] and puff decaying [(c) and (d)] for $\Rey=2046$.  In each panel, we plot $q$ or $u$ for every 5 time steps. In order to avoid overlaps of these configurations in a single panel, we shift each configuration along $y$-axis when time is incremented. More precisely, we plot $q^t_x+(1/5) t$ or $u^t_x+(1/5) t$ for several $t$ ($t=0, 1,2,\dots$) in each panel. We re-define $t=0$ as the time a few hundred steps before the puff splitting or the puff decaying. The true starting times of these simulations are more than $10^4$ steps before this $t=0$. 
} 
\end{center}
\end{figure*}

We set the parameters ($d,\epsilon_1,\epsilon_2,c,\gamma,\beta$) to ($0.15,0.04,0.2,0.45,0.95,0.4$) according to Ref.~\cite{PhysRevE.84.016309}. In the main text, we only change the value of the parameter $\Rey$ without modifying the others.  
We start a simulation from a localized configuration, such as the Kronecker-delta configuration with a randomly chosen intensity between 0 and 1. After an initial relaxation time, the puff dynamics becomes statistically stable (especially for $\Rey \sim 2040$). 
In Fig~\ref{fig:f_explanation}(b,c), we plot snapshots of a puff configuration. Although these dynamics are stable, one can sometimes observe splitting and decaying of puffs in a long-time simulation. 
The snapshots in Fig.~\ref{fig:f_split} demonstrate such splitting and decaying, observed after simulating the system around $10^5$ steps.
The duration of time before the splitting and the decaying is determined stochastically following an exponential law (see Fig.~12 in Ref.~\cite{PhysRevE.84.016309} for the observation of this law within this model, and also see Refs.~\cite{hof2006finite,PhysRevLett.101.214501,de_Lozar589,avila2010transient,kuik2010quantitative} in more realistic settings). 
%

%
% Figure dynamics
% Figure dynamics
%

\subsection{Total turbulence intensity}

We define a total turbulence intensity $\lambda$ as
\begin{equation}
\lambda(X) = \sum_{x=0}^{L} q_x.
\label{eq:defReactionCoordinate}
\end{equation}
We show a typical time series of $\lambda(X)$ for splitting and decaying in Fig.~\ref{fig:timeseries} of the main text. 
From the figure, we find that $\lambda(X)$ does not take a value less than 1 when there is at least one puff, but it takes less than 1 after the puff decays. We thus define  
\begin{equation}
\lambda_{\rm decay} = 1
\end{equation}
as a threshold value of the lower bound of $\lambda(X)$, below which the puff completely decays. 
At the same time, $\lambda(X)$ takes a value around 40 when double puffs occur, and 
it takes (almost) always a value less than 40 in a presence of a single puff.  Since we focus on
the dynamics of a single puff and its decaying, we thus define 
\begin{equation}
\lambda_{\rm split} = 41
\end{equation}
as a threshold value for the upper bound of $\lambda(X)$~\cite{caption1}.

\section{Super-exponential fitting to the probability distribution function $p(\lambda)$}
\label{Append_NumericalStudy}

Here we show a super-exponential fitting to the bulk part of the probability distribution function $p(\lambda)$.

\begin{figure}
\begin{center}
\includegraphics[width=0.45\textwidth]{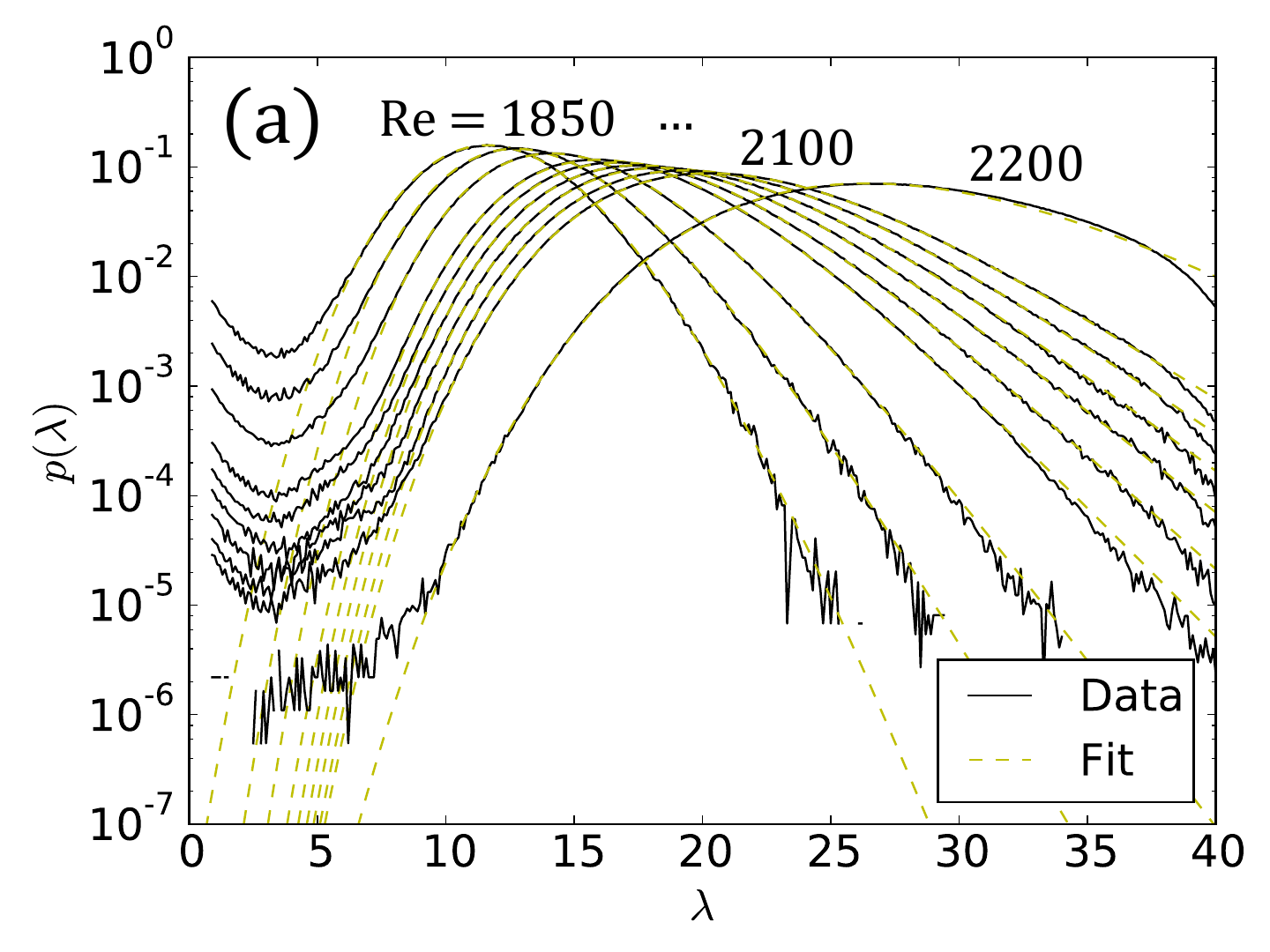} 
\includegraphics[width=0.45\textwidth]{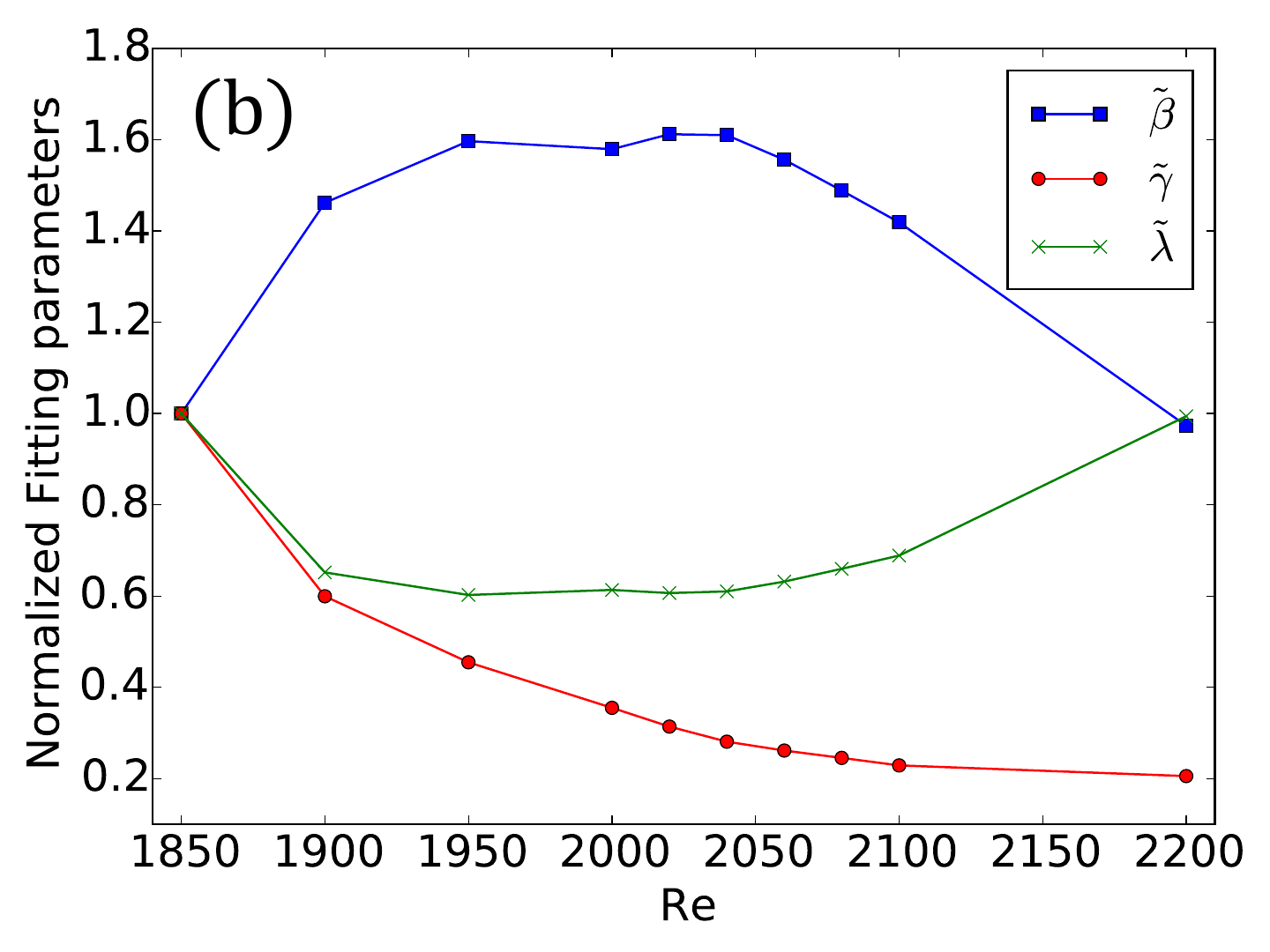} 
\caption{\label{fig:plambda_withFitting} {\bf (a)} The probability distribution functions of $\lambda$, $p(\lambda)$ for several values of the Reynolds number $\Rey$ (Re = 1850, 1900, 1950, 2000, 2020, 2040, 2060, 2080, 2100, 2200), which are measured from brute-force simulations.  
We also plot the super-exponential curve (\ref{eq:def_fitting}) by using the parameters determined by fitting to the data. The agreement between the fitting curve and the numerical data is excellent for a certain range of $\lambda$. {\bf (b)} The (normalized) fitting parameters, $\tilde \beta/\tilde \beta(1850)$, $\tilde \gamma/\tilde \gamma(1850)$, $\tilde \lambda/\tilde \lambda(1850)$ as a function of the Reynolds number, where the values of $\tilde \beta(1850)$, $\tilde \gamma(1850)$, $\tilde \lambda(1850)$ are 0.1012, 1.574, 38.75, respectively. One can see that $\tilde \beta$ and $\tilde \lambda$ show a plateau in the range from  $\Rey = 1950$ to $\Rey = 2050$, whereas $\tilde \gamma$ is monotonically decreasing.} 
\end{center}
\end{figure}

We consider a probability distribution function $p(\lambda)$ 
defined as a derivative of the accumulative probability $P(\lambda)$, (\ref{lambda_}).
We show in Fig.~\ref{fig:plambda_withFitting} numerical examples of $p(\lambda)$ for several Reynolds numbers, together with the derivative of the fitting function (\ref{eq:fitexp}):
\begin{equation}
p_{\rm fit}(\lambda) = C \rm \exp \left \{ - \exp \left [ - \tilde \beta \left (\lambda -  \tilde \lambda \right ) \right ] - \tilde \gamma (\lambda - \tilde \lambda) \right \},
\label{eq:def_fitting}
\end{equation}
where $C$ is a normalization constant, and $\tilde \beta, \tilde \gamma, \tilde \lambda$ are fitting parameters. We note that this fitting function reduces to a Gumbel distribution function~\cite{gumbel1935valeurs} when  $\tilde \beta =  \tilde \gamma$. Interestingly, as shown in Fig.~\ref{fig:plambda_withFitting}(a), the fitting curve describes perfectly the numerical data in a certain range of $\lambda$ for several different Reynolds numbers. We also plot the (normalized) fitting parameters, $\tilde \beta/\tilde \beta(1850)$, $\tilde \gamma/\tilde \gamma(1850)$, $\tilde \lambda/\tilde \lambda(1850)$ in Fig~\ref{fig:plambda_withFitting}(b). The data indicate $\beta \neq \gamma$ in general, namely the distribution function is not described by Gumbel distribution. 

To provide an insight into this super-exponential form (\ref{eq:def_fitting}), we introduce an effective Brownian motion describing typical dynamics of $\lambda(X)$.
Since it has been observed that the puff-decaying time scale is simply described by a memoryless exponential law \cite{hof2006finite,PhysRevLett.101.214501,de_Lozar589,avila2010transient,kuik2010quantitative}, we assume that the typical dynamics of $\lambda(X)$ itself can be described by the following Brownian process $\lambda^t_{\rm s}$
\begin{equation}
\frac{d \lambda^t_{\rm s}}{dt} = f(\lambda^t_{\rm s}) + \xi^t,
\end{equation}
where $\xi^t$ is a Gaussian white noise satisfying zero mean $\langle \xi^t \rangle = 0$ and the delta-function correlation $\langle \xi^t \xi^s \rangle = D \delta(t-s)$ with a noise intensity $D$. The function $f(\lambda)$ represents the
effective force describing the dynamics of the turbulence intensity. For this function, we consider two contributions, $f_{\rm -}(\lambda)$ and $f_{\rm +}(\lambda)$. The first contribution is to reduce the size of the puff at the interface between the turbulent region and the Laminar region. This contribution does not depend on the value of $\lambda$, so that we model this effect as a constant term $f_0$, {\it i.e., } $f_{\rm -}(\lambda)=-f_0$. The second contribution is to enlarge the turbulent region. When the turbulence intensity is small, puffs immediately develop their intensity, whereas when the turbulence intensity is large,
the dynamics immediately lose such a driving force.  
To model this behavior, we assume that $f_{+}(\lambda)$ is written as an exponential function $f_{+}(\lambda) =  \alpha e^{ -\beta (\lambda - \lambda_0)} $ with three parameters $\alpha, \beta$ and $\lambda_0$. 
To sum up $f_+$ and $f_-$, we get $f(\lambda) = - f_0 +  \alpha e^{ - \beta (\lambda - \lambda_0)} $. Since the stationary distribution function of $\lambda_{\rm s} ^ t$, $p_{\rm st}(\lambda)$, is derived as the canonical distribution function $e^{(1/D) \int d\lambda f (\lambda)}$, we thus obtain
\begin{equation}
p_{\rm st}(\lambda) = \tilde C \exp \left \{  - \frac{f_0}{D} \lambda - \frac{\alpha}{D\beta} \exp \left [- \beta (\lambda - \lambda_0) \right ] \right \}
\label{eq:p_st}
\end{equation}
with a normalization constant $\tilde C$. By redefining the parameters in this expression, one can see that $p_{\rm st}(\lambda)$ is equivalent to the fitting function  (\ref{eq:def_fitting}). 
%

%%%%%%%%%%%%%%%%%%%%%%
\begin{table}
\begin{center}
          \caption{\label{Table:Parametersused_ndecay} $n_{\rm decay}^{\rm max}$: the values of $n_{\rm decay}$ when we stop the measurement procedures (i-iii) in Section~\ref{Sec:Fluctuations_TimeScale}. For Re-control method, $n_{\rm decay}^{\rm max}$ for $\lambda_1$ and $\Rey_0$ specified in Table~\ref{Table:Parametersused} is shown.}
          \begin{tabular}{c||c|c}
\vphantom{\Big|}                                             
          $\Rey_1$                     &   Brute-force  & Re-control  
\\[1mm]  \hline  \hline \vphantom{\Big|}  
 1800  &  3600  &  -- 
\\ [1mm]  \hline \vphantom{\Big|}  
1850     &  3600 	 & --    
\\[1mm]   \hline  \vphantom{\Big|}  
1900     & 3600    & -- 
\\[1mm]  \hline  \vphantom{\Big|}  
 1950  &  3600  &   -- 
\\ [1mm]  \hline \vphantom{\Big|}  
2000     &  3600 	 & --    
\\[1mm]   \hline  \vphantom{\Big|}  
2050     & 3600    & --  
\\[1mm]  \hline  \vphantom{\Big|}  
2100  &  3600  & 3600
\\ [1mm]  \hline \vphantom{\Big|}  
2150     & 3600 & 3600   
\\[1mm]   \hline  \vphantom{\Big|}  
2200     & 3600   & 3600  
\\[1mm]  \hline  \vphantom{\Big|}  
2250     & 2746    & 3600  
\\[1mm]  \hline  \phantom{\Big|}  
 2300  &  578  & 3600 
\\ [1mm]  \hline \vphantom{\Big|}  
2350     &  -- 	 & 1647    
\\[1mm]   \hline  \vphantom{\Big|}  
2400     & --    & 151 
\\[1mm]   \hline  \vphantom{\Big|}  
2450     & --    & 149  
\\[1mm]   \hline  \vphantom{\Big|}  
2500     & --    & 12
\\[1mm] \hline
          \end{tabular}  
        \end{center}
 \end{table}
%%%%%%%%%%%%%%%%%%%%%%

\section{Values of $n_{\rm decay}$ when stopping the measurements}
\label{parameters_used_ndecay}

For getting the data in Fig.~\ref{fig:Ptail}, \ref{fig:Td}, \ref{fig:Timescale_log_large}, we stop the measurement procedure (i-iii) in Section~\ref{Sec:Fluctuations_TimeScale} when
$n_{\rm decay}$ reaches a certain value, which we denote by $n_{\rm decay}^{\rm max}$. 
We summarize $n_{\rm decay}^{\rm max}$ in Table~\ref{Table:Parametersused_ndecay}.

\section{Values of parameters $\lambda_0$, $\lambda_1$, $\Rey_0$ for Fig.~\ref{fig:Timescale_log_large}}
\label{parameters_used}

In Table~\ref{Table:Parametersused}, we summarize the parameters used in Fig.~\ref{fig:Timescale_log_large} for Re-control method. These parameters are chosen according to the condition in Table~\ref{Table:Parameter_setting}. 

For $\Rey=2500$, in order to observe puff decaying event in our limited simulation time, we needed to set $\lambda_1$ to be close enough to $\lambda_{\rm ms}^{\Rey_1}$. This is
a possible reason why the predicted value of puff-decaying time scale for $\Rey=2500$ in Fig.~\ref{fig:Timescale_log_large} is 
slightly higher than the supper-exponential curve, because as seen from Fig.~\ref{fig:Td}(a),
as $\lambda_1$ gets close to $\lambda_{\rm ms}^{\Rey_1}$, the method becomes much faster,
but the estimated value
of $T_d$ tends to be larger than the correct value.

%%%%%%%%%%%%%%%%%%%%%%
%\begin{table*}
%\begin{center}
%          \caption{\label{Table:Parametersused} Parameters $\lambda_0$, $\lambda_1$, $\Rey_0$ used in Fig.~%\ref{fig:Timescale_log_large}}
%          \begin{tabular}{c||c|c|c|c|c|c|c|c|c}
%\vphantom{\Big|}                                             
%          $\Rey_1$                     &   2100  & 2150  & 2200   & 2250  & 2300 & 2350 & 2400 & 2450 & 2500
%\\[1mm]  \hline  \hline \vphantom{\Big|}  
%$\lambda_0 (= \bar \lambda_{\Rey_1})$  &  21.75  & 24.55 & 27.50  & 30.23 & 32.45 & 34.22 & 35.65 & 36.81 & %37.73
%\\ [1mm]  \hline \vphantom{\Big|}  
%$\lambda_1$                            &  14 	 & 16    & 18     & 18    & 22    & 22    & 24    & 24    & 24 
%\\[1mm]   \hline  \vphantom{\Big|}  
%$\Rey_1 - \Rey_0$                       & 200    & 200  & 200   & 200  & 200  & 200  & 200  & 300  & 300  
%\\[1mm] \hline
%          \end{tabular}  
%        \end{center}
% \end{table*}
%%%%%%%%%%%%%%%%%%%%%%%

%%%%%%%%%%%%%%%%%%%%%%
\begin{table}
\begin{center}
          \caption{\label{Table:Parametersused} The values of $\lambda_0$, $\lambda_1$, $\Rey_0$ used in Fig.~\ref{fig:Timescale_log_large} for Re-control method}
          \begin{tabular}{c||c|c|c}
\vphantom{\Big|}                                             
$\Rey_1$ &   $\lambda_0 (= \bar \lambda_{\Rey_1})$  & $\lambda_1$   & $\Rey_1 - \Rey_0$
\\[1mm]  \hline  \hline \vphantom{\Big|}  
  2100  &  21.75  & 14 & 200   
\\ [1mm]  \hline \vphantom{\Big|}  
  2150  &  24.55  & 16  & 200     
\\[1mm]   \hline  \vphantom{\Big|}  
  2200  &  27.50  & 18  & 200   
\\[1mm]   \hline  \vphantom{\Big|}  
  2250   & 30.23  & 18  & 200   
\\[1mm]   \hline  \vphantom{\Big|}  
  2300   & 32.45  & 22  & 200   
\\[1mm]   \hline  \vphantom{\Big|}  
  2350   & 34.22  & 22  & 200   
\\[1mm]   \hline  \vphantom{\Big|}  
  2400   & 35.65  & 24  & 200   
\\[1mm]   \hline  \vphantom{\Big|}  
  2450   & 36.81  & 24  & 300   
\\[1mm]   \hline  \vphantom{\Big|}  
  2500   & 37.73  & 24  & 300   
\\[1mm] \hline
          \end{tabular}  
        \end{center}
 \end{table}
%%%%%%%%%%%%%%%%%%%%%%

%\Urlmuskip=0mu plus 2mu\relax
\bibliographystyle{plain_url}
\bibliography{draft.bib}

\end{document}